\pdfoutput=1
\documentclass[11pt,twoside,a4paper,cmspaper,final,collab]{cms-tdr}
\def\svnVersion{5e75c48}\def\svnDate{2026/03/20}\def\cmsCernNoTag{CERN-EP-2025-193}\def\cmsCernDate{\today}\def\cmsMessage{Published in Physical Review D as \href{http://dx.doi.org/10.1103/c4z7-tqvc}{\doi{10.1103/c4z7-tqvc}.}}
\begin{document}\cmsNoteHeader{JME-24-001}

\newlength\cmsTabSkip\setlength{\cmsTabSkip}{1ex}
\providecommand{\cmsTable}[1]{\resizebox{\textwidth}{!}{#1}}

\newcommand{\qt}{\ensuremath{q_{\mathrm{T}}}\xspace}
\newcommand{\uparal}{\ensuremath{u_{\parallel}}\xspace}
\newcommand{\uperp}{\ensuremath{u_{\perp}}\xspace}
\newcommand{\vqt}{\ensuremath{\vec{q}_{\mathrm{T}}}\xspace}
\newcommand{\vut}{\ensuremath{\vec{u}_{\mathrm{T}}}\xspace}
\newcommand{\KERAS} {{\textsc{Keras}}\xspace}
\newcommand{\TENSORFLOW} {{\textsc{TensorFlow}}\xspace}
\newcommand{\ADAMW}{{\textsc{AdamW}}\xspace}
\newcommand{\DeepMET}{{\textsc{DeepMET}}\xspace}
\ifthenelse{\boolean{cms@external}}{\providecommand{\cmsLeft}{upper\xspace}}{\providecommand{\cmsLeft}{left\xspace}}
\ifthenelse{\boolean{cms@external}}{\providecommand{\cmsRight}{lower\xspace}}{\providecommand{\cmsRight}{right\xspace}}

\cmsNoteHeader{JME-24-001}

\title{\DeepMET: Improving missing transverse momentum estimation with a deep neural network}

\date{\today}

\abstract{
At hadron colliders, the net transverse momentum of particles that do not interact with the detector (missing transverse momentum, \ptvecmiss) is a crucial observable in many analyses.   In the standard model, \ptvecmiss originates from neutrinos. Many beyond-the-standard-model particles, such as dark matter candidates, are also expected to leave the experimental apparatus undetected.  This paper presents a novel deep neural network based \ptvecmiss estimator, \DeepMET, developed by the CMS Collaboration at the LHC. The \DeepMET algorithm produces a weight for each reconstructed particle based on its properties. The estimator is based on the negative vector sum of the weighted transverse momenta of all reconstructed particles in an event. Compared with other estimators currently employed by CMS, \DeepMET improves the \ptvecmiss resolution by 10--30\%, shows improvement for a wide range of final states, is easier to train, and is more resilient against the effects of additional proton-proton interactions accompanying the collision of interest.
}

\hypersetup{
    pdfauthor={CMS Collaboration},
    pdftitle={DeepMET: Improving missing transverse momentum estimation with a deep neural network},
    pdfsubject={CMS},
    pdfkeywords={CMS, machine learning, missing transverse momentum}
}

\maketitle

\section{\label{sec:intro}Introduction}
At the CERN LHC, experiments, such as CMS~\cite{Chatrchyan:2008zzk} and ATLAS~\cite{ATLAS:2008xda}, use reconstruction algorithms that estimate the four-momentum, production vertex, and type of all final-state particles to extract the maximum information from the high energy proton-proton ($\Pp\Pp$) collision data. The reconstructed particles consist of muons, electrons, photons, as well as charged and neutral hadrons, which are then clustered into jets that are the experimental signatures of energetic quarks, gluons, and tau leptons.
The presence of particles that do not interact with the detector, such as standard model (SM) neutrinos or new electrically neutral, weakly interacting particles, \eg, dark matter particles, can only be inferred indirectly, by measuring the so-called missing transverse momentum, denoted as \ptvecmiss, with magnitude \ptmiss. At a hadron collider, the component transverse to the beam direction of the net momentum of the initial-state partons is close to zero. The negative of the vector sum of all reconstructed particles' transverse momenta therefore estimates \ptvecmiss:
\begin{equation}\label{eq:ptmiss} 
\ptvecmiss = -\sum_{i \in \text{all reco.}}\vec{p}_{\mathrm{T},i}.
\end{equation}
The \ptvecmiss stemming from weakly interacting particles will be collectively referred to as ``genuine \ptvecmiss" in what follows. 
The accuracy of the \ptvecmiss estimate is affected by the particle reconstruction efficiencies and \pt resolutions, detector finite coverage and noise, and the ability of the reconstruction algorithm to include particles from the leading $\Pp\Pp$ interaction, while rejecting the potentially large number of low-momentum particles from additional $\Pp\Pp$ interactions in the same or nearby bunch crossings (pileup).
These factors collectively make it challenging to measure \ptvecmiss with high accuracy and resolution. Pileup mitigation algorithms have been adopted and optimized in CMS that are effective in reducing this contribution to \ptvecmiss resolution~\cite{CMS:2020ebo,Bertolini:2014bba,CMS-DP-2021-001}.

The complexity of the \ptvecmiss measurement suggests that  modern machine-learning (ML) techniques could yield a more accurate estimator.
The CMS Collaboration has done previous work showing the utility of this approach.  An estimator that is based on multivariate analysis (MVA) technique using boosted decision trees, referred to as MVA-\ptvecmiss~\cite{CMS:2014xym}, is employed in CMS Run-1 analyses and demonstrates a relative improvement in precision of around 20\% compared with a more traditional calculation~\cite{CMS:2014wdm}, particularly for the $\PH\to\PGt\PGt$ process.
However, because MVA-\ptvecmiss was trained on specific physics processes and their global event properties, it does not, in general, have optimal performance for other physics processes and kinematic regions.

This paper introduces a new algorithm \DeepMET based on a deep neural network (DNN). For each reconstructed particle in the event, the DNN produces a unique weight based on the properties of the particles. The \DeepMET \ptvecmiss estimator is based on the negative of the vector sum of the weighted transverse momenta of all particles in an event. 
In this paper, we describe \DeepMET and compare its performance to that of  other \ptvecmiss estimators. Our performance metrics are based on the response and resolution for  the $x$- and $y$-components of a \ptvecmiss estimator, the components parallel and perpendicular to the transverse direction of a \PZ boson  (for samples enriched in this process), or the absolute value of \ptmiss. The response is defined as the ratio of the measured average to the true average. The response can be smaller than unity due to \eg, particles below reconstruction thresholds or the response of the calorimeters. It can be larger due to resolution effects, noise, and pileup. The resolution can be calculated both with and without correction for the response. Since it can be difficult to know the true \ptvecmiss in an event, artificial \ptvecmiss can be induced in events with little to no genuine \ptvecmiss by excluding leptons from the \ptvecmiss calculation (\eg, the muons in a $\PZ \to \PGm\PGm$ event) and comparing with the corresponding precisely measured quantity calculated from the leptons. 

Compared with conventional \ptvecmiss estimators, \DeepMET shows a relative improvement in the \ptvecmiss resolution of 10--30\%, and is more resilient to the effects of pileup. 
Due to the choice of DNN architecture and its inputs, \ie, the usage of reconstructed particles and the exclusion of event-level observables, the algorithm has limited information about the event topology and is trained only with typical $\Pp\Pp$ events.
As a consequence, \DeepMET is trained quickly and generalizes well, showing similarly strong performance in both collision data and a diverse set of simulated physics processes.

The paper is structured as follows. Section~\ref{sec:cmsdetector} gives an overview of the CMS detector. In Section~\ref{sec:related}, a summary of existing algorithms is presented, along with their strengths and weaknesses. Sections~\ref{sec:data} and~\ref{sec:selections} contain details of the data sets and the object and event selections. After a brief introduction to the \ptvecmiss parametrization, Section~\ref{sec:algo} details the algorithm, its inputs and outputs, and the associated loss functions. Performance metrics and comparisons with alternative algorithms are discussed in Section~\ref{sec:perf}. A method to calibrate simulation performance to align with real collision data is outlined in Section~\ref{sec:calib}. Section~\ref{sec:deeper} explores the source of the enhanced performance of \DeepMET. Benchmarks for the computational needs are given in Section~\ref{sec:implementation}.
Finally, a summary is presented in Section~\ref{sec:summary}.

Tabulated results are provided in the HEPData record for the performance comparisons in this analysis~\cite{hepdata}.

\section{\label{sec:cmsdetector}The CMS detector}
The central feature of the CMS apparatus is a superconducting solenoid of 6\unit{m} internal diameter, providing a magnetic field of 3.8\unit{T}. Within the solenoid volume are a silicon pixel and strip tracker, a lead tungstate crystal electromagnetic calorimeter (ECAL), and a brass and scintillator hadron calorimeter (HCAL), each composed of a barrel and two endcap sections. Forward hadron (HF) calorimeters extend the pseudorapidity ($\eta$) coverage to high values ($|\eta|\leq5$) assuring very good hermeticity. Muons are detected in gas-ionization detectors embedded in the steel flux-return yoke outside the solenoid. A more detailed description of the CMS detector, together with a definition of the coordinate system used and the relevant kinematic variables, is reported in Ref.~\cite{Chatrchyan:2008zzk}.

Events of interest are selected using a two-tiered trigger system. The first level, composed of custom hardware processors, uses information from the calorimeters and muon detectors to select events of interest at a rate of around 100\unit{kHz} within a fixed time interval of about 4\mus~\cite{CMS:2020cmk}. The second level, known as the high-level trigger,
consists of a farm of processors running a version of the full event reconstruction
software optimized for fast processing, and reduces the event rate to $\mathcal{O}$(1\unit{kHz})
before data storage~\cite{Khachatryan:2016bia,CMS:2024aqx}.

\section{\label{sec:related}Reconstruction, pileup mitigation, and machine learning}
\subsection{Particle-flow reconstruction}

Primary collision vertices (PV) corresponding to $\Pp\Pp$ collisions are found using reconstructed  charged particles~\cite{CMS:2014pgm}.
The primary $\Pp\Pp$ interaction vertex corresponding to the hardest scattering in the event, evaluated using tracking information alone, is referred to as the leading primary vertex (LPV), as described in Section 9.4.1 of Ref.~\cite{CMS-TDR-15-02}. Other PVs are referred to as pileup vertices.

A particle-flow (PF) algorithm~\cite{CMS:2017yfk} aims to reconstruct and identify each individual particle in an event, using an optimized combination of information from the various elements of the CMS detector. The energy of a photon is obtained from the ECAL measurement. The energy of an electron is determined from a combination of the electron momentum at the primary interaction vertex as determined by the tracker, the energy of the corresponding ECAL cluster, and the energy sum of all bremsstrahlung photons spatially compatible with originating from the electron track. The \pt of a muon is obtained from the curvature of the corresponding track. The energy of a charged hadron is determined from a combination of the momentum measured in the tracker and the matching ECAL and HCAL energy deposits, corrected for the response function of the calorimeters to hadronic showers. Finally, the energy of a neutral hadron is obtained from the corresponding corrected ECAL and HCAL energies. The reconstructed particles from the PF algorithm are referred to as PF candidates.

Muons are measured in the range $\abs{\eta} < 2.4$, with detection planes made using three technologies: drift tubes, cathode strip chambers, and resistive-plate chambers. The single-muon trigger efficiency exceeds 90\% over the full $\eta$ range, and the efficiency to reconstruct and identify muons is greater than 96\%. Matching muons to tracks measured in the silicon tracker results in a relative \pt resolution, for muons with \pt up to 100\GeV, of 1\% in the barrel and 3\% in the endcaps. Measurements made with cosmic ray muons show that, in the central region of the detector, the \pt resolution is better than 7\% for muons with \pt up to 1\TeV~\cite{CMS:2018rym}.

For each event, hadronic jets are clustered from these reconstructed particles using the infrared and collinear safe anti-\kt algorithm~\cite{Cacciari:2008gp, Cacciari:2011ma} with a distance parameter of 0.4. Jet momentum is determined as the vector sum of all particle momenta in the jet, and is found from simulation to be, on average, within 5 to 10\% of the true momentum over the whole \pt spectrum and detector acceptance.
Jet energy corrections are derived from simulation studies so that the average measured energy of jets becomes identical to that of particle level jets. In situ measurements of the momentum balance in dijet, photon+jet, {\PZ}+jet, and multijet events are used to determine any residual differences between the jet energy scale (JES) in data and in simulation, and appropriate corrections are made. The jet energy resolution (JER) typically amounts to 15--20\% at 30\GeV, 10\% at 100\GeV, and 5\% at 1\TeV~\cite{CMS:2016lmd}. Additional selection criteria are applied to each jet to remove jets potentially dominated by instrumental effects or reconstruction failures.

The reconstruction also produces a \ptvecmiss estimator, PF \ptvecmiss, which is calculated using all reconstructed PF candidates in an event~\cite{CMS:2019ctu} and is further corrected for the JES of reconstructed jets. The \ptvecmiss relies on the accurate measurement of the reconstructed physics objects, namely muons, electrons, photons, hadronically decaying taus, jets, and unclustered energy ($E_U$). The $E_U$ is the contribution from the PF candidates not associated with any of the previous physics
objects.

\subsection{Pileup mitigation}

Pileup affects jet and \ptvecmiss measurements, since contributions from additional tracks and calorimetric energy depositions increase the apparent jet momentum and change the subsequent \ptvecmiss estimation. To mitigate such effects on jets,  charged-hadron subtraction (CHS) removes tracks identified as originating from pileup vertices from jets~\cite{CMS:2020ebo}. Corrections are then applied to remove on average the remaining contributions from neutral particles. This correction to the \pt of the jets is then propagated to the \ptvecmiss estimation~\cite{CMS:2016lmd,CMS:2019ctu}.

The pileup-per-particle identification algorithm (PUPPI)~\cite{CMS:2020ebo,Bertolini:2014bba,CMS-DP-2021-001} can additionally be used to mitigate the effect of pileup  by making use of local shape information, event pileup properties, and tracking information.  Charged particles identified as originating from pileup vertices are discarded. For each neutral particle, a local shape variable, that distinguishes collinear energy deposits originating from the hard scatter from soft/diffuse energy deposits attributable to particles originating from pileup interactions, is defined using the surrounding charged particles that are compatible with the LPV within the tracker acceptance ($\abs{\eta} < 2.5$) and using all particles in the region outside of the tracker coverage. The momenta of the neutral particles are then rescaled according to their probability to originate from the LPV as deduced from the local shape variable. We denote this probability as the ``PUPPI weight'' in the following.  The rescaling supersedes the need for jet-based offset corrections~\cite{CMS:2020ebo}. The corresponding \ptvecmiss is referred to as PUPPI \ptvecmiss. In the studies of this paper, ``PUPPI v15" tune is used~\cite{CMS-DP-2021-001}. 

The CHS and PUPPI algorithms have been shown to improve the precision of \ptvecmiss measurements significantly~\cite{CMS:2019ctu}. However, the PUPPI algorithm requires extensive tuning. Its efficacy depends both on the physics process and kinematic region, and it is less effective in low-pileup data. Consequently, its \ptvecmiss resolution is not always optimal.

\section{\label{sec:data}Data and simulated samples}
This paper uses a data sample collected by the CMS experiment at a center-of-mass energy of 13\TeV during 2016. The data set corresponds to an integrated luminosity of $\lumi=16.8\fbinv$, with an average pileup of 23 additional $\Pp\Pp$ interactions~\cite{CMS:2021xjt}. The \DeepMET algorithm was deployed for the remainder of Run 2 and Run 3 as well, where it showed comparable performance gains.

Several Monte Carlo (MC) event generators are used to simulate the {\PZ}+jets and \ttbar samples for the ML training and {\PW}+jets samples for the post-training performance studies. The {\PV}+jets ($\PV=\PW, \PZ$) samples are simulated at next-to-leading order (NLO) in perturbative quantum chromodynamics (QCD) with the \MGvATNLO~2.3.3 event generator~\cite{Alwall:2014hca}. The \ttbar samples are simulated with \POWHEG~2.0~\cite{Nason:2004rx,Frixione:2007vw,Alioli:2008gx,Alioli:2010xd}. The \PYTHIA~8.230~\cite{Sjostrand:2014zea} package is used for parton showering, hadronization, and the underlying event simulation, with the CP5 tune~\cite{Skands:2014pea, Sirunyan:2019dfx}. The NNPDF~3.1~\cite{NNPDF:2017mvq} set of parton distribution functions (PDFs) at next-to-NLO (NNLO) in QCD is used as the default.

Additional sets of simulated event samples are used to understand how the performance of the \DeepMET reconstruction depends on the physical process. The processes span a variety of final states, including Higgs bosons, top quarks, heavy-flavor jets, \PGt leptons, and beyond-the-SM dark-matter candidates.
These samples are produced using the same PDFs with the same tunes.
Four types of Higgs boson processes are generated with \POWHEG at NLO in perturbative QCD:  Higgs boson production via vector boson fusion with the Higgs boson decaying to invisible particles ($\PH\to\text{invisible}$), Higgs boson pair production via gluon fusion with one Higgs boson decaying to a pair of \PGt leptons (\PGt decay inclusively) and the other Higgs boson decaying to a pair of \PQb quarks ($\PH\PH\to\PQb\PQb\PGt\PGt$), and Higgs bosons produced in association with a \ttbar pair followed by the Higgs boson either decaying to a pair of \PQb quarks or to a pair of muons and the top quarks decaying to leptons.
Furthermore, two sets corresponding to the production of particles using the Simplified Model Spectra (SMS) framework of supersymmetry are considered~\cite{Alwall:2008ag,LHCNewPhysicsWorkingGroup:2011mji}, both with $R$ parity conservation and a lightest supersymmetric particle, a neutralino \PSGczDo that acts as a dark-matter candidate.
One set consists of top squark-antisquark pair production with a top squark mass of 350\GeV and a \PSGczDo mass of 335\GeV.  The final state containing two \PQb quarks, four light quarks, and two \PSGczDo particles~\cite{CMS:2019zmd} is selected ($\PSQt\to\PQb\PQq\PAQq\PSGczDo$, denoted SMS T2tt).  The second set is pair production of two of the heavier neutralinos with mass 400\GeV, with each neutralino decaying into a \PZ boson and a \PSGczDo with a mass of 1\GeV~\cite{CMS:2022sfi} (denoted  SMS TChiZZ).
Both of these processes are generated with \MGvATNLO in version 2.3.3 at LO.
An overview of the simulated samples and their respective final states is given in Table~\ref{tab:samples_mc_performance}.

\begin{table*}[htb]
	\topcaption{Overview of the simulated event samples used for performance studies.}
    \centering
    \cmsTable{
	\begin{scotch}{lll}
        Process & Production mode, decay chain & Final-state objects \\
        \hline
        \textit{Standard Model} \\
        $\PZ$ boson & $\PZ$+jets, $\PZ \to \ell^+ \ell^-$ & Charged leptons, jets\\
        $\PW$ boson & $\PW$+jets, $\PW \to \ell \nu$ & Charged leptons, jets, prompt neutrinos\\
        Top quark pair & $\ttbar\to2\ell2\nu2\PQb$ &  Muons, \PQb quark jets, prompt neutrinos\\
        \textit{Higgs Physics} \\
        Invisible Higgs & $\PQq\PQq\PH$, $\PH \to \PZ\PZ^* \to 4\nu$ & Forward jets, prompt neutrinos \\
        Di-Higgs & $\Pg\Pg \to \PH\PH\to\PQb\PQb\PGt\PGt$ & $\tau$ leptons, \PQb quark jets,  prompt neutrinos \\
        $\PQt\PQt\PH (\PQb\PQb)$ & $\PQt\PQt\PH$, $\PH \to \PQb \PAQb$ & Jets, \PQb quark jets, charged leptons, prompt neutrinos \\
        $\PQt\PQt\PH (\PGm\PGm)$ & $\PQt\PQt\PH$, $\PH \to \mu^+ \mu^-$ & Muons, jets, \PQb quark jets, prompt neutrinos \\
        \textit{Supersymmetry} \\
        Top squark pair & $\Pg\Pg\to\PSQt\,\PASQt\to 2\PQb 4\PQq 2\PSGczDo$ & Jets, \PQb quark jets, dark matter candidates \\
        Heavy neutralino & $\Pg\Pg\to\PSGczDt\PSGczDt\to\PZ\PZ\PSGczDo\PSGczDo $ & Charged leptons, jets, dark matter candidates \\
        
	\end{scotch}
    }
    \label{tab:samples_mc_performance}
\end{table*}

The simulation of the interactions of all final-state particles with the CMS detector is done with \GEANTfour~\cite{Agostinelli:2002hh}. The simulated events are reconstructed using the same algorithms used for the data. The simulated events include pileup, with the number of additional $\Pp\Pp$ interactions matching that observed in data~\cite{Lange:2015sba,CMS:2018mlc}.

\section{Physics object and event selections}\label{sec:selections}

To select {\PW}+jets, {\PZ}+jets, and \ttbar processes, candidate events are required to pass the isolated single-muon trigger that requires \pt above 24\GeV.

The selected muon candidates are required to satisfy a set of quality criteria based on the number of spatial measurements in the silicon tracker and the muon system, as well as the fit quality of the combined muon track~\cite{CMS:2018rym}. Muon candidates are required to have $\pt>25\GeV$, $\abs{\eta}<2.4$, and pass the so-called ``tight" identification and isolation selections~\cite{CMS:2018rym}.

The {\PZ}+jets and \ttbar events used in the algorithm training described in Section~\ref{sec:algo} are required to have two muons with opposite charge passing the above selections. In the studies quantifying the algorithm performance described in Section~\ref{sec:zmumu}, the dimuon invariant mass is also required to satisfy $80<m(\PGm\PGm)<100\GeV$ in order to select $\PZ\to\PGm\PGm$ events and veto those with genuine \ptvecmiss (mainly from \PGt leptons or electroweak decays inside the heavy-flavor jets). 

The {\PW}+jets simulated events and data used in Section~\ref{sec:wjets} are required to have one muon candidate passing the muon selections. 

For the studies presented in Section~\ref{sec:perf_other} validating the \ptvecmiss performance in other simulated physics processes,
no object or event selection is applied.  

\section{\label{sec:algo}The \DeepMET algorithm}
 The DNN operates on each PF candidate individually.  The candidate's inputs include eight continuous and three categorical features:

\begin{itemize}
\item Continuous features:
 \begin{itemize}
   \item $d_{xy}$, the impact parameter with respect to the LPV in the transverse plane for charged PF candidates, zero for neutral PF candidates~\cite{CMS:2014pgm};
   \item $d_z$, the impact parameter with respect to the LPV along the beam direction for charged PF candidates, zero for neutral PF candidates;
   \item the PUPPI weight;
   \item five kinematic features per particle: $\eta$, mass, $p_{x}$, $p_{y}$, and \pt;
 \end{itemize}
 
\item Categorical features:
 \begin{itemize}
  \item Particle ID: a unique integer number for the seven different types of PF candidates, \ie, electrons, muons, photons, charged hadrons, neutral hadrons, and hadronic and electromagnetic PF candidates in the HF;
  \item charge of the particle ($-1$, 0, or $+1$);
  \item an LPV association flag, referred to as the \texttt{fromLPV} flag. This flag indicates one of four possible associations between the LPV and a charged PF candidate: used in the LPV fit; used in a pileup vertex fit; not used in any of the PV fits, but closest to one of the pileup vertices in the $z$ direction; and not used in any of the PV fits but closest to the LPV in the $z$ direction.
  \end{itemize}
\end{itemize} 
 
 The output of the DNN is a weight ($w_{i}$) and two bias terms ($b_{i,x}, b_{i,y}$). 
The \DeepMET estimator output is the negative of the weighted \ptvec sum of all PF candidates together with their bias contributions.  Its $x$ and $y$ components are given by:
\begin{equation}\label{eq:dmetcal}
\begin{aligned}
 p^{\text{miss}}_{x}&=-\sum_{i} (w_{i}p_{i,x}+b_{i,x}),
 p^{\text{miss}}_{y}&=-\sum_{i} (w_{i}p_{i,y}+b_{i,y}).
\end{aligned}
\end{equation}
This calculation is similar to the PUPPI \ptvecmiss calculation. The two bias terms in the model were initially intended to enable distinct adjustments in the $x$ and $y$ directions independently and to correct measurement biases. However, in practice it is found that they have minimal impact on the performance.

\subsection{The DNN architecture}
\begin{figure*}[htp]
   \centering
    \includegraphics[width=0.99\textwidth]{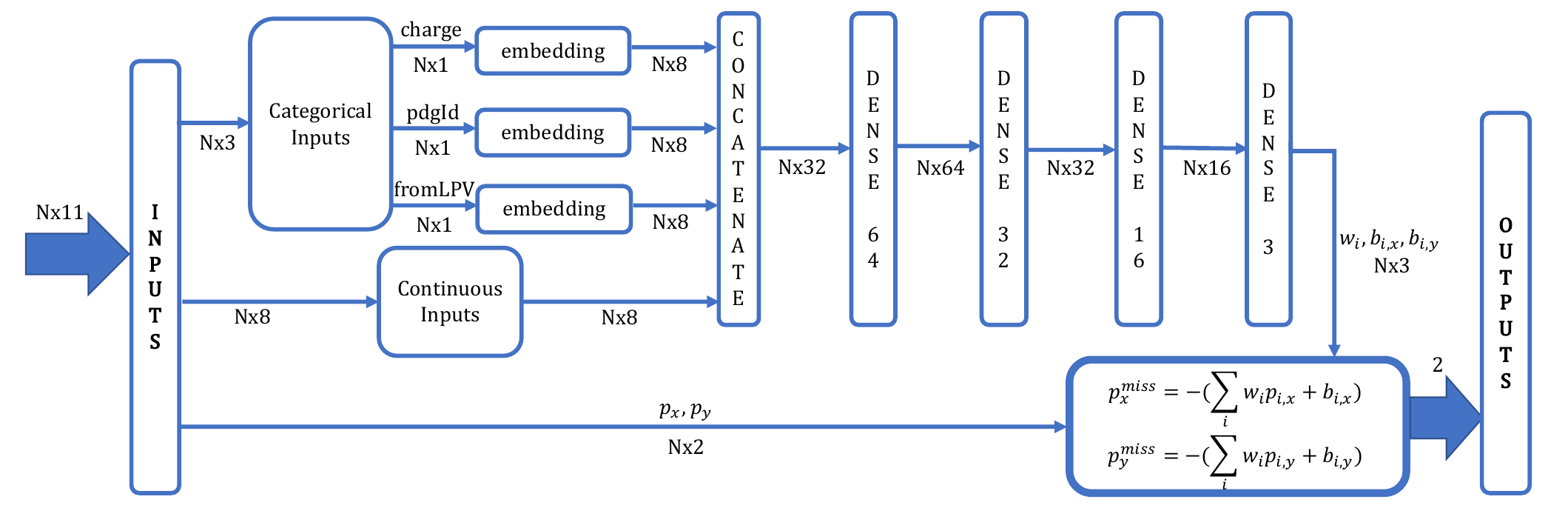}
    \caption{\label{fig:dnnarch}The \DeepMET DNN architecture. For each event, all PF candidates in the event are considered as input. $N\times n$ represents the number of PF candidates in the event multiplied by the dimensionality of the per-particle feature space, where $n$ can be 1, 2, 3, 8, 16, 32, and 64 in the architecture.
    }
\end{figure*}

Figure~\ref{fig:dnnarch} shows the DNN architecture. The inputs are the 11 features of each input PF candidate. After the input layer, each of the three categorical features will go through one embedding layer and become an 8-dimensional tensor, so that the feature space can be learned and the differences can be represented by distances in the embedded space.

The three 8-dimensional tensors from the embedding layers are then concatenated together with the eight continuous features, and passed to a few dense (or fully-connected) layers with progressively reducing numbers of nodes. These layers act only on each particle individually and can also be considered as
one-dimensional convolutions. After each dense layer, batch normalization~\cite{bn} is applied (not shown in Fig.~\ref{fig:dnnarch} for simplicity) to reduce the covariate shifts and make the training converge faster. The outputs of these dense layers are the weights $w_i$, and the biases $b_{i,x}$ and $b_{i,y}$ of all the input PF candidates. In the last layer, the event \ptvecmiss is calculated following Eq.~\ref{eq:dmetcal}.

The total number of trainable parameters in the \DeepMET neural network is 4541.
Such a small set of trainable parameters allows the training to converge quickly, and also significantly reduces the inference time. Studies have also been conducted using more layers and larger number of nodes per layer, and the effect on the performance was within a few percent. 

The DNN architecture is implemented with \KERAS~\cite{chollet2015keras}, which acts as an interface for the \TENSORFLOW~\cite{tensorflow} library. The \ADAMW optimizer~\cite{Loshchilov:2017bsp} is used for gradient optimizations. 

\subsection{Training}

The DNN of \DeepMET is trained with the {\PZ}+jets and \ttbar simulated samples discussed in Section~\ref{sec:data}. Events are required to pass the dimuon selections described in Section~\ref{sec:selections}. Using two samples allows a wide range of jet multiplicities and particle \pt distributions: {\PZ}+jets events tend to have particles with low \pt; the \ttbar events tend to have high \pt and typically contain high-\pt neutrinos.

The \ptvecmiss performance in {\PV}+jets processes is measured by comparing the momentum of the vector boson $\vqt$ to that of the hadronic recoil system $\vut$ (defined as the vector \pt sum of all PF candidates except for the decay products of the vector boson)~\cite{CMS:2019ctu}. The vector boson momenta $\vqt$ are measured as the vector \pt sum of the two muons for \PZ bosons, or taken directly from the generator-level information in simulations of \PW bosons.

The components of the hadronic recoil parallel and perpendicular to the boson momentum vector are denoted by \uparal and \uperp, respectively. These are used to study the \ptvecmiss response and resolution. The response of \ptvecmiss is defined as $-\langle\uparal\rangle/\langle\qt\rangle$; the resolutions of \uparal and \uperp are calculated using the root-mean-square of $\uparal+\qt$ and $\uperp$ distributions, and are denoted $\sigma(\uparal)$ and $\sigma(\uperp)$, respectively~\cite{CMS:2019ctu}.

Training is performed on an \textsc{NVIDIA} Quadro P6000 graphics card using a few tens of thousands of events from the {\PZ}+jets sample and approximately the same number of events from the \ttbar sample. The batch size is set to 128, and the training and validation splitting is set to 4:1. The training takes about 30~minutes to converge to stable results.  

Further studies are performed by scanning the mixture fractions of the two samples and the hyperparameters (such as the number and size of dense layers, learning rate, training and validation splitting, and the batch size), and by increasing the number of total events in the training and validation samples. \DeepMET performance is found to be only weakly sensitive to modifications of these (hyper)parameters. 

\subsection{Loss function}

The loss function used in the training is the mean squared error loss:
\begin{equation}
L=\frac{1}{2}[(p^\text{miss}_{x}-\hat{p}^\text{miss}_{x})^{2}+(p^\text{miss}_{y}-\hat{p}^\text{miss}_{y})^{2}]
\end{equation}
where $p^\text{miss}_{x,y}$, $\hat{p}^\text{miss}_{x,y}$ are the predicted and generator-level components of \ptvecmiss in the $x,y$ directions.

\section{\label{sec:perf}The \DeepMET performance}
The \DeepMET performance in data and MC simulations is discussed in this section. Improvements in resolutions are observed, with the response closer to unity, and more pileup resilience compared with PF \ptvecmiss and PUPPI \ptvecmiss.

The following performance studies focus primarily on the muon channel; however, comparable performance improvements were observed in the electron channel.

\subsection{Performance on \texorpdfstring {{\PZ}+jets}{Z+jets} events}\label{sec:zmumu}
\begin{figure}[htp]
  \centering
  \includegraphics[width=0.49\textwidth]{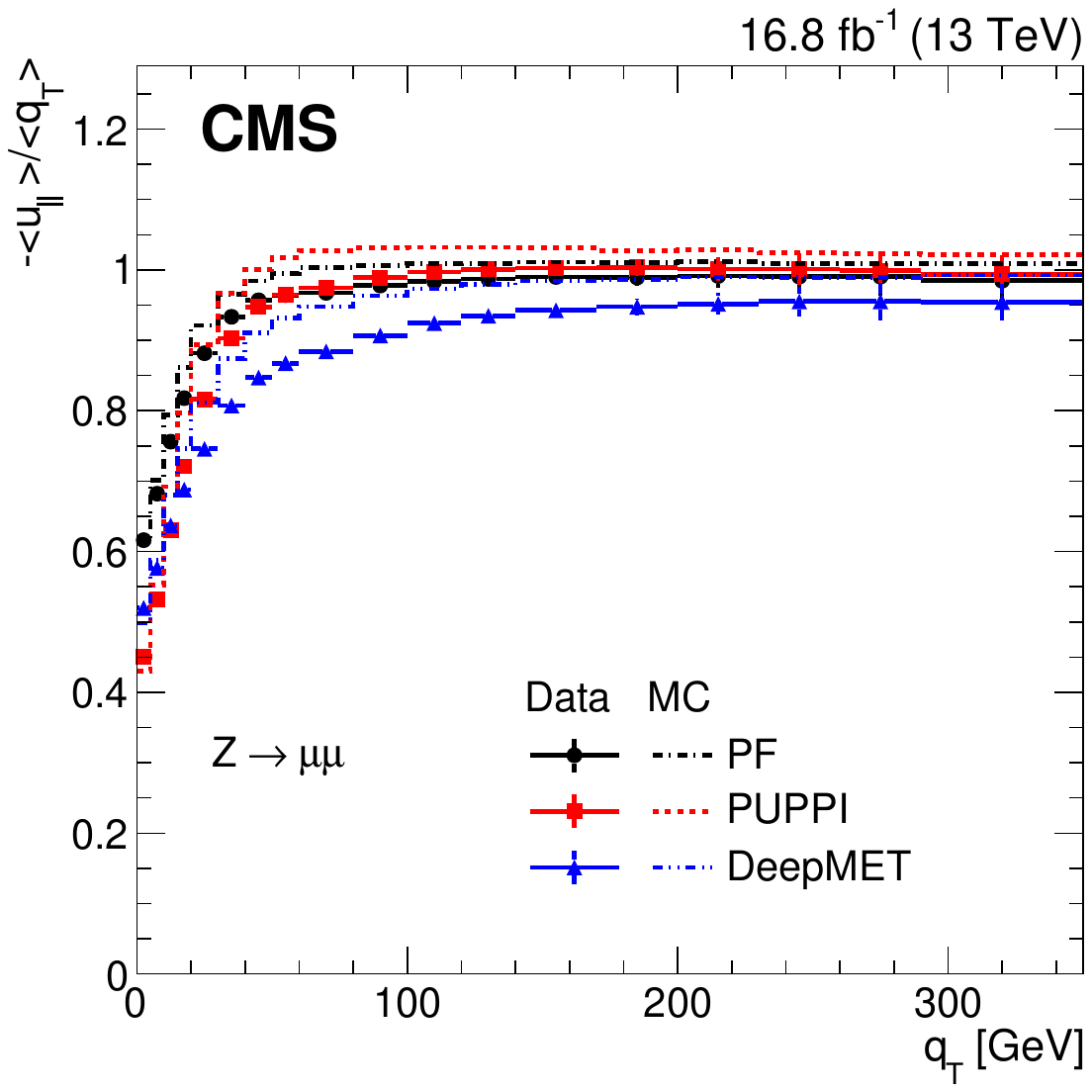}
  \caption{\label{fig:deepresponse_zdata}Recoil responses of different \ptvecmiss estimators in data (markers) and MC simulations (dashed) after the $\PZ\to\mu\mu$ selections.}
\end{figure}

The \ptvecmiss performance is studied using the events passing the {\PZ}+jets selections in Section~\ref{sec:selections}. Such events should have little or no genuine \ptvecmiss. The responses of different \ptvecmiss estimators are shown in Fig.~\ref{fig:deepresponse_zdata}. Markers correspond to collision data, and dashed lines indicate MC simulations. The \DeepMET algorithm maintains a good, albeit 5-10\% worse, response than the PF and PUPPI \ptvecmiss estimators. 

\begin{figure}[htp]
  \centering
  \includegraphics[width=0.49\textwidth]{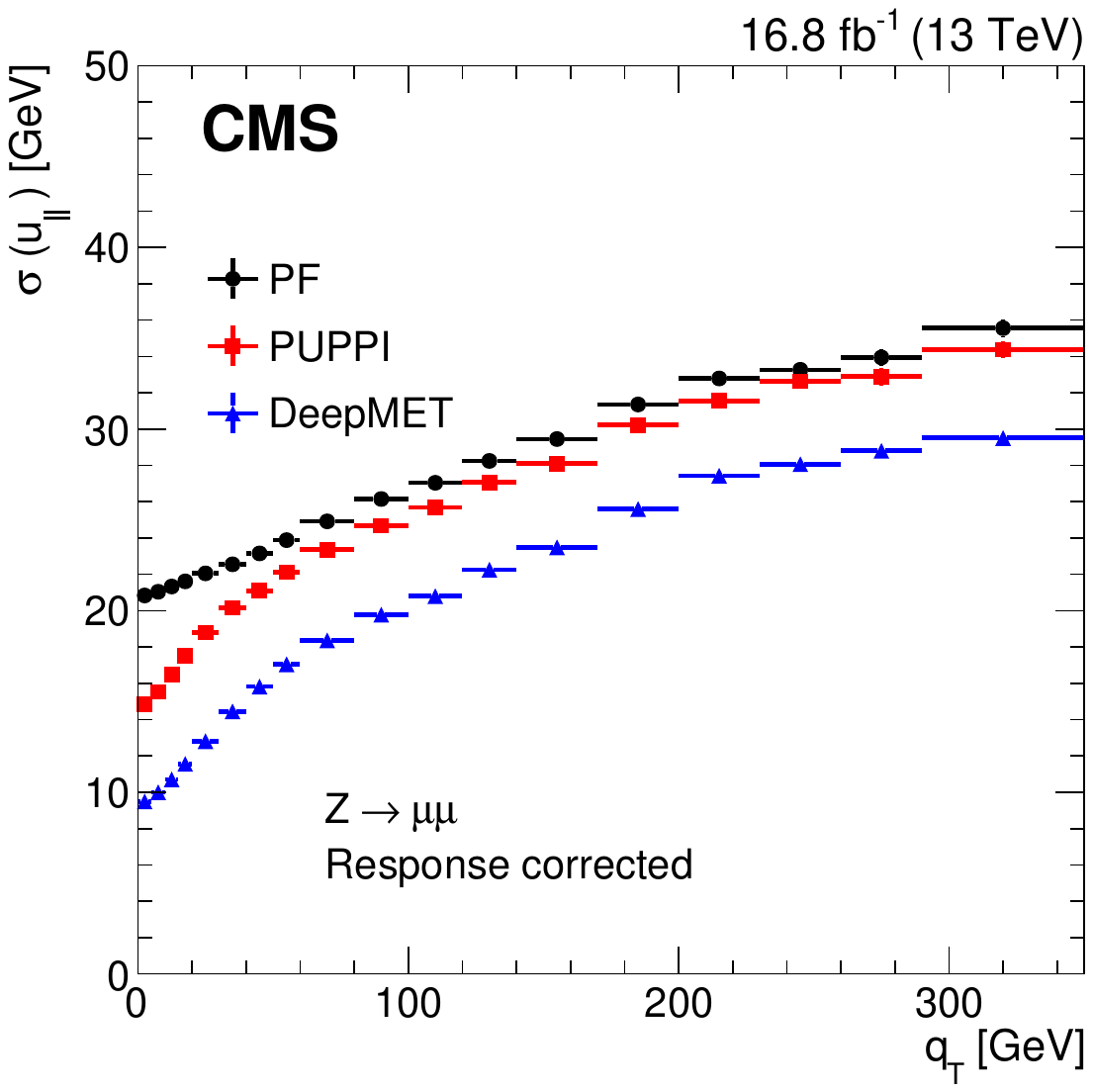}
  \includegraphics[width=0.49\textwidth]{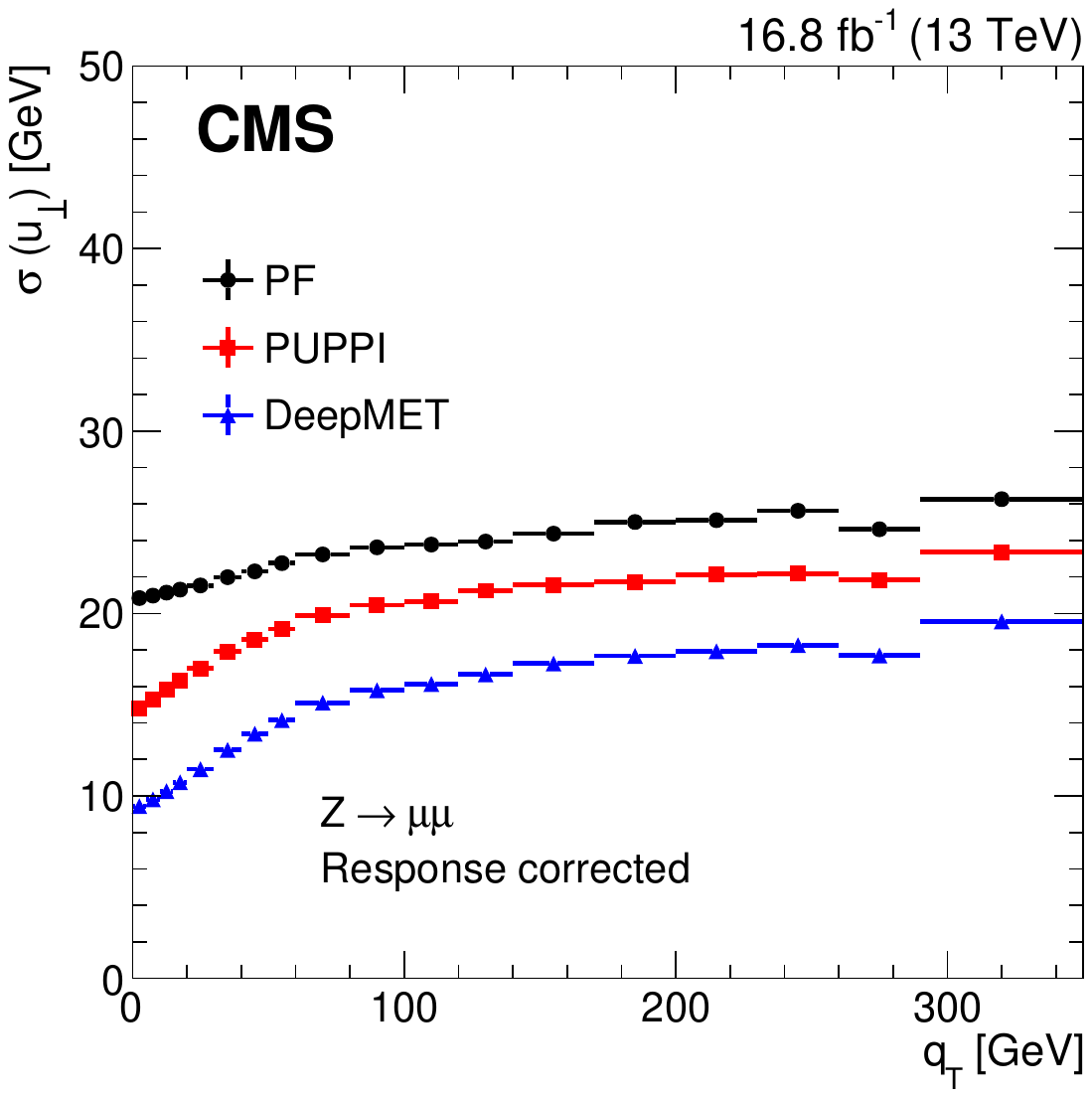}
  \caption{\label{fig:deepresol_zdata_vs_qt_scaled}Response-corrected resolutions of \uparal (\cmsLeft) and \uperp (\cmsRight) \vs \qt of different \ptvecmiss estimators in data after the $\PZ\to\mu\mu$ selections.
  }
\end{figure}

\begin{figure}[htp]
  \centering
  \includegraphics[width=0.49\textwidth]{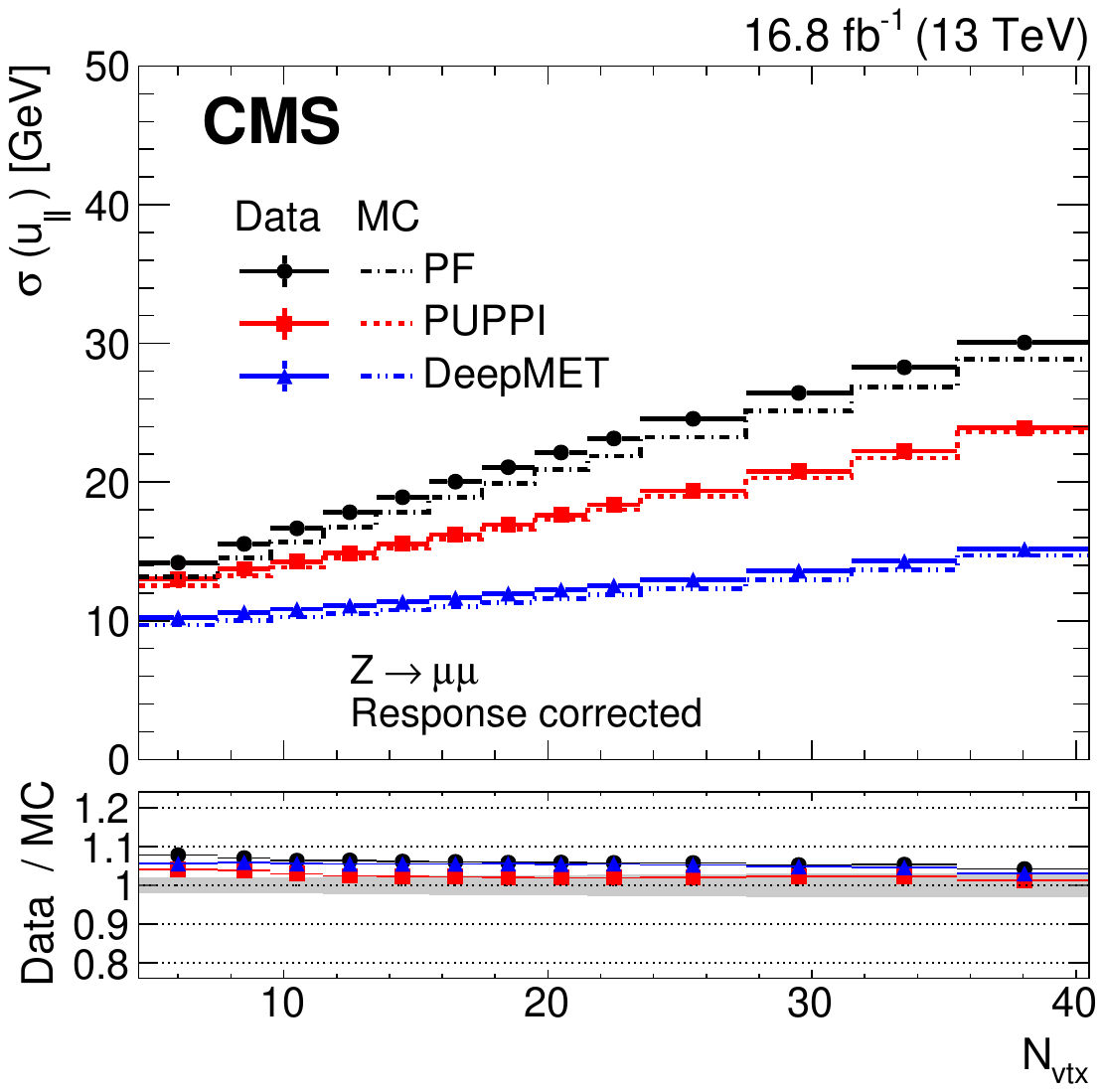}
  \includegraphics[width=0.49\textwidth]{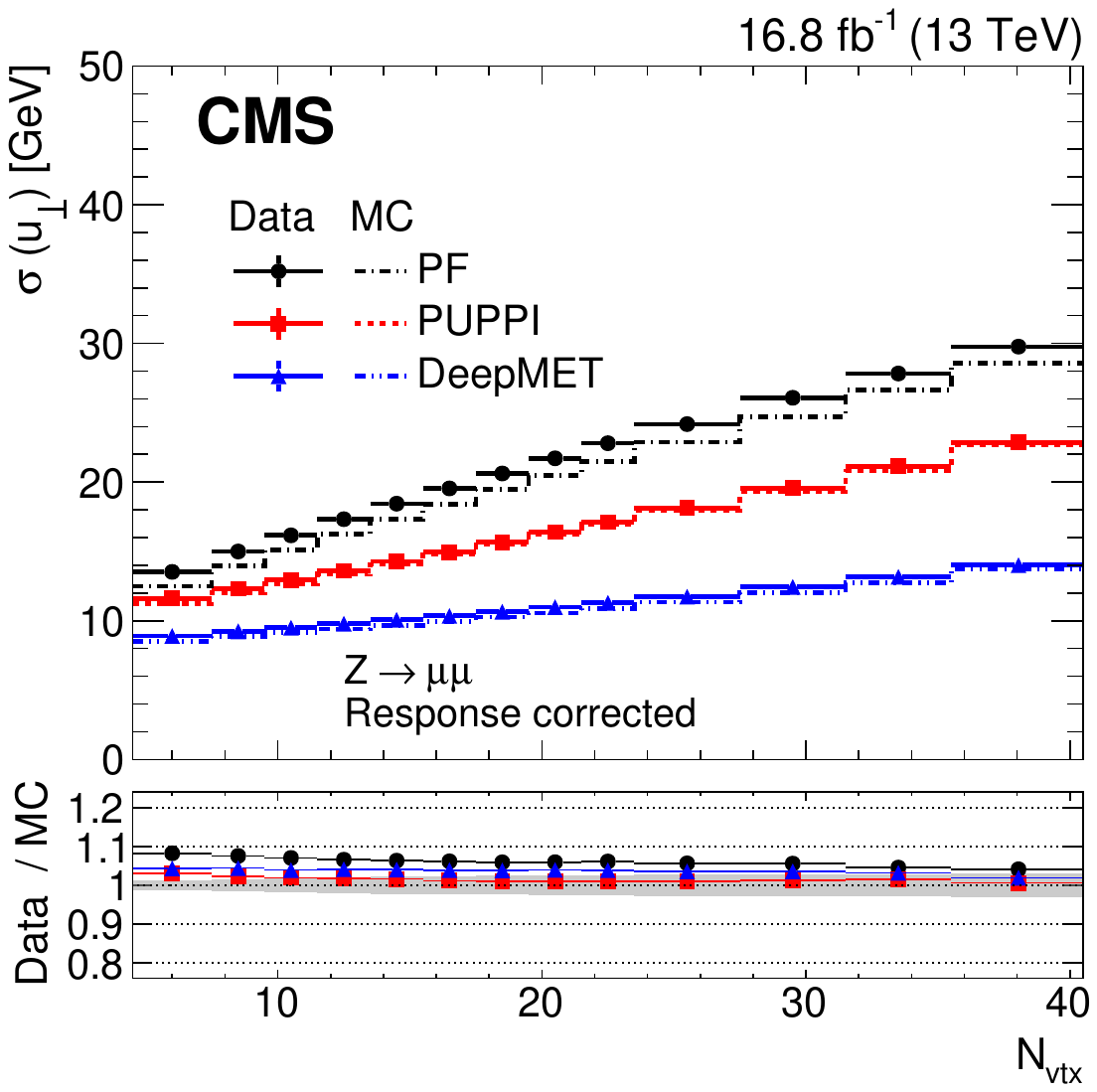}
  \caption{\label{fig:deepresol_zdata_scaled}Response-corrected resolutions of \uparal (\cmsLeft) and \uperp (\cmsRight) \vs number of reconstructed PVs of different \ptvecmiss estimators in data (solid) and MC simulations (dashed) 
  after the $\PZ\to\mu\mu$ selections. The systematic uncertainties for PUPPI \ptvecmiss due to the JES, the JER, and $E_U$ are added in quadrature and displayed with the gray band.
  }
\end{figure}

To properly account for the response effect on the resolutions, for each \ptvecmiss estimator, the average response $-\langle\uparal\rangle/\langle\qt\rangle$ is calculated using the events in the response plateau region (\ie, $\qt>150\GeV$), and the resolutions are scaled by the inverse of this average response, referred to as ``response-corrected resolutions". The response-corrected resolutions of \uparal and \uperp \vs \qt and the number of reconstructed PVs are shown in Figs.~\ref{fig:deepresol_zdata_vs_qt_scaled} and~\ref{fig:deepresol_zdata_scaled}, respectively.
The \DeepMET algorithm maintains good resolutions and pileup resilience. Despite slightly worse response than the PF and PUPPI \ptvecmiss estimators, \DeepMET resolutions are  better even after including the response corrections. For example, compared with PUPPI \ptvecmiss, the \DeepMET shows a 2-10\% worse response but achieves a 20-30\% resolution gain around 20 reconstructed primary vertices, significantly outperforming PUPPI in this regime.

\begin{figure}[htp]
  \centering
  \includegraphics[width=0.49\textwidth]{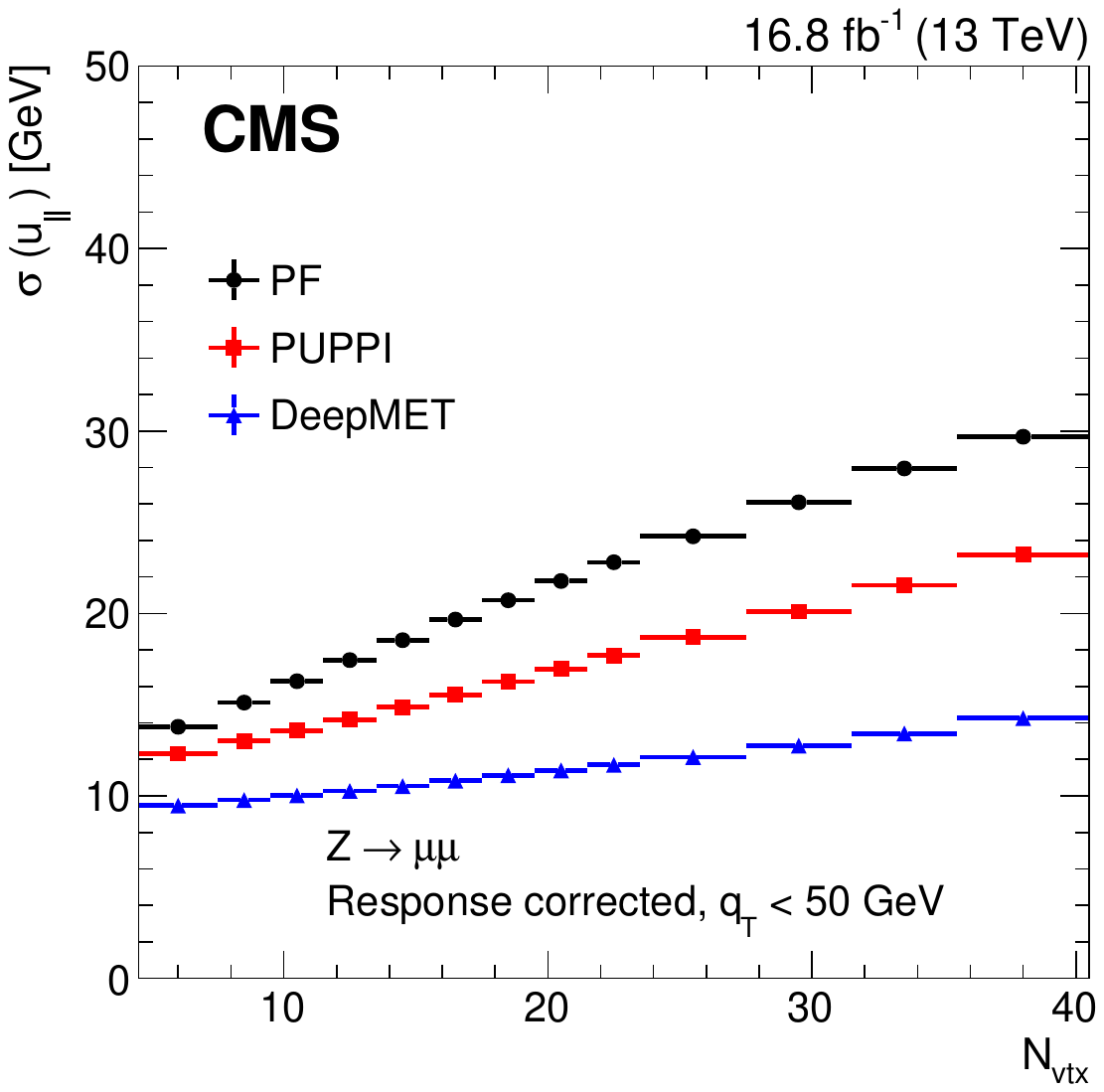}
  \includegraphics[width=0.49\textwidth]{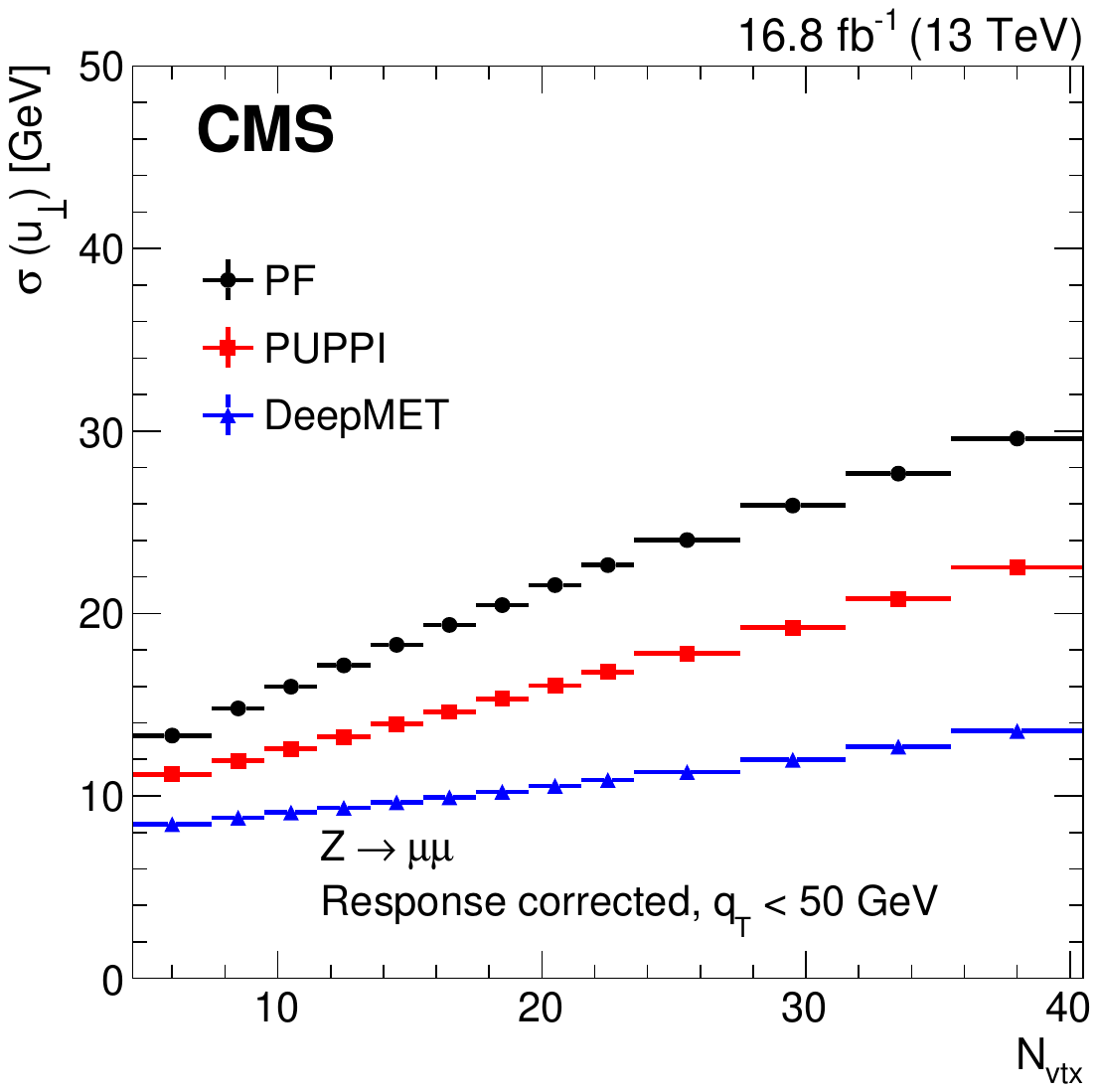}
  \caption{\label{fig:deepresol_zdata_scaled_lowqt}Response-corrected resolutions of \uparal (\cmsLeft) and \uperp (\cmsRight) \vs number of reconstructed PVs of different \ptvecmiss estimators in data
  after the $\PZ\to\PGm\PGm$ selections in the region with $\qt$ smaller than 50\GeV.
  }
\end{figure}

\begin{figure}[htp]
  \centering
  \includegraphics[width=0.49\textwidth]{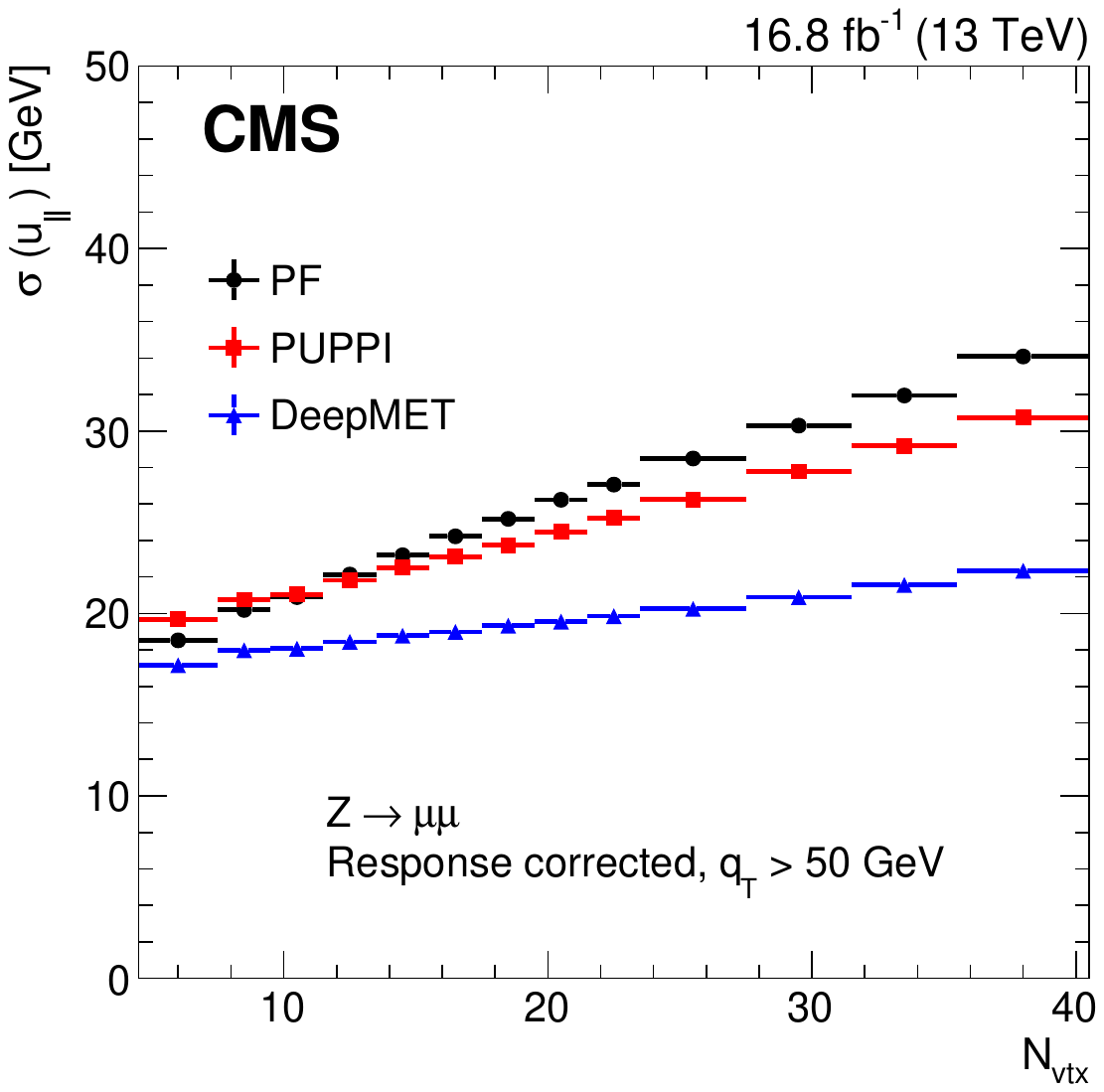}
  \includegraphics[width=0.49\textwidth]{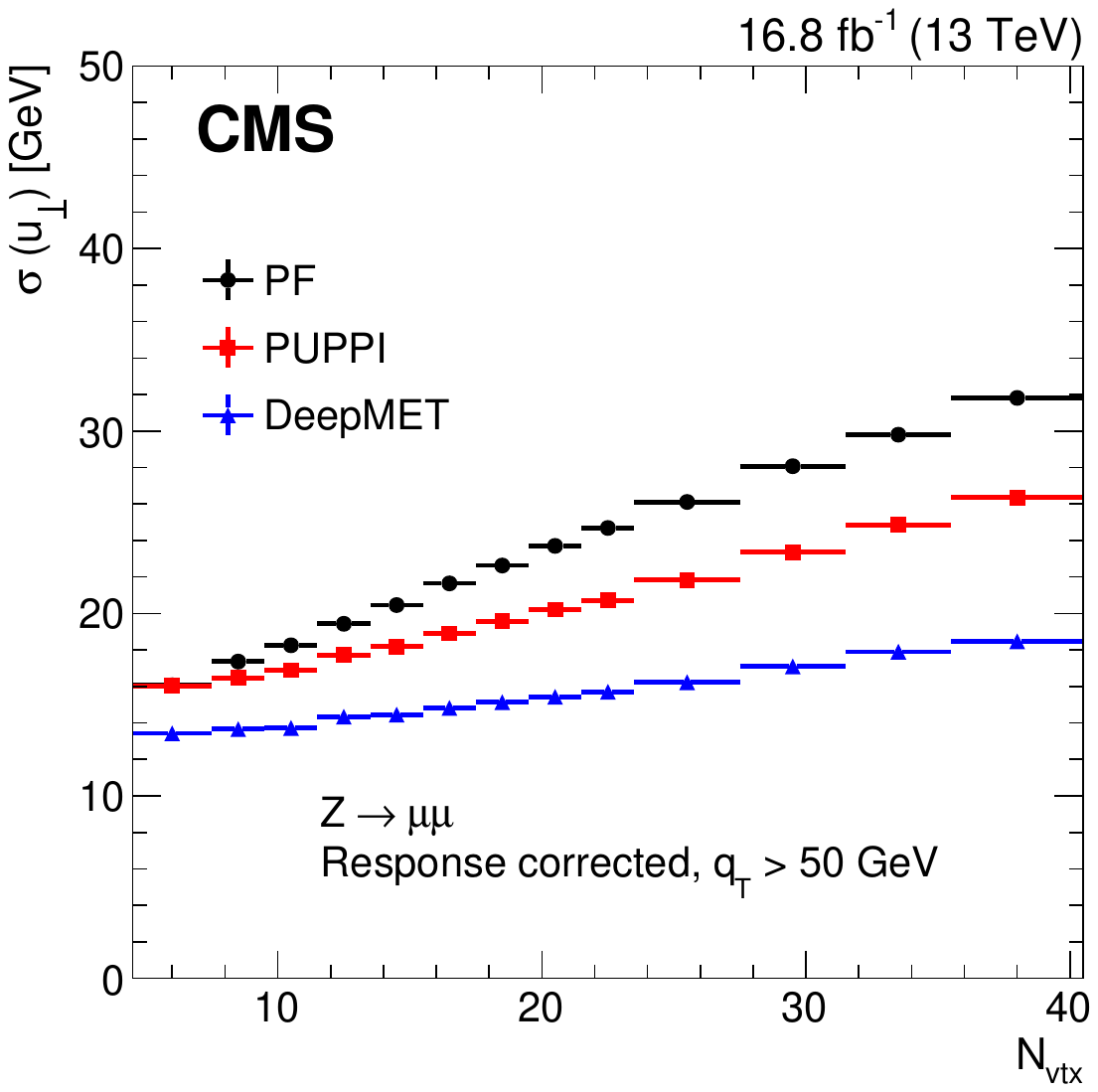}
  \caption{\label{fig:deepresol_zdata_scaled_highqt}Response-corrected resolutions of \uparal (\cmsLeft) and \uperp (\cmsRight) \vs number of reconstructed PVs of different \ptvecmiss estimators in data
  after the $\PZ\to\PGm\PGm$ selections in the region with $\qt$ larger than 50\GeV.
  }
\end{figure}

The comparisons of response-corrected resolutions of different \ptvecmiss estimators in $\PZ\to\PGm\PGm$ events for the low- and high-\qt regions are provided in Figs.~\ref{fig:deepresol_zdata_scaled_lowqt} and~\ref{fig:deepresol_zdata_scaled_highqt}, respectively, corresponding to the region where $\qt<(>)\, 50\GeV$. In both regions, PUPPI \ptvecmiss is better than PF \ptvecmiss, and \DeepMET outperforms both \ptvecmiss estimators.

\subsection{The PV-Agnostic \DeepMET and performance on \texorpdfstring{{\PW}+jets}{W+jets} events}\label{sec:wjets}

The performance has also been studied in {\PW}+jets MC samples, in which the \PW boson decays into a muon and a(n) (anti)neutrino. Typically, this signal process involves a single isolated muon, genuine \ptvecmiss, and several soft tracks as the \pt distribution of \PW bosons is low with the most probable value around 5\GeV~\cite{CMS:2016mwa}. In this study, the accuracy of LPV identification in approximately 1\% of the events in the {\PW}+jets simulations is compromised, where the LPV is either reconstructed but incorrectly identified as a pileup vertex, or is not reconstructed at all. This occurs more frequently in events with muons at high $\abs{\eta}$, where the quality of the muon track is worse. Such mistakes in the LPV identification can result in incorrect PV-track association flags and PUPPI weights, which in turn introduce inaccuracies in subsequent calculations of PUPPI and \DeepMET \ptvecmiss. 

To handle such scenarios for high-precision $\PW(\PGm\PGn)$ analyses, a PV-Agnostic version of \DeepMET is  developed, which is robust to the choice of the LPV from the PVs~\cite{CMS:2024lrd}. In this version of the algorithm, the default LPV selection is discarded, and instead the PV closest in $z$ to the leading muon with $dz(\mathrm{PV}, \PGm)<0.2\cm$ is chosen. If there is no PV satisfying the requirement (\ie, the LPV is not successfully reconstructed), a ``pseudo"-LPV is built, using ($x$, $y$) of the beam spot (luminous region of the $\Pp\Pp$ collisions~\cite{CMS:2014pgm}), and the leading muon $z$ position. In the \DeepMET inputs, $d_z$'s and the PUPPI weights are re-computed with respect to the new LPV, while \verb|fromLPV| and $d_{xy}$ are not used, since they cannot be recalculated precisely. The model is retrained using the reduced set of inputs.

\begin{figure*}[htp]
\centering
\includegraphics[width=0.49\textwidth]{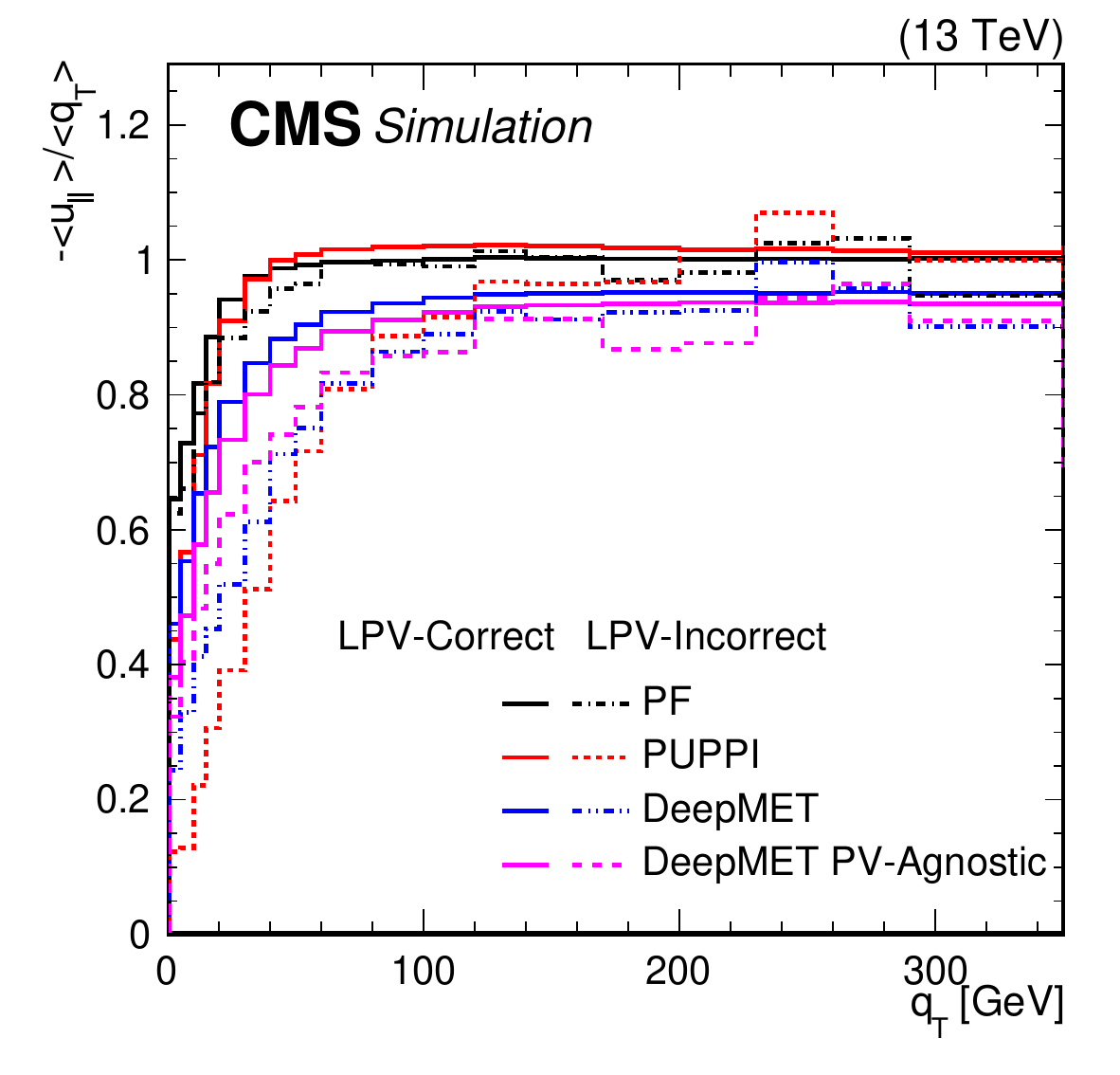}\\
\includegraphics[width=0.49\textwidth]{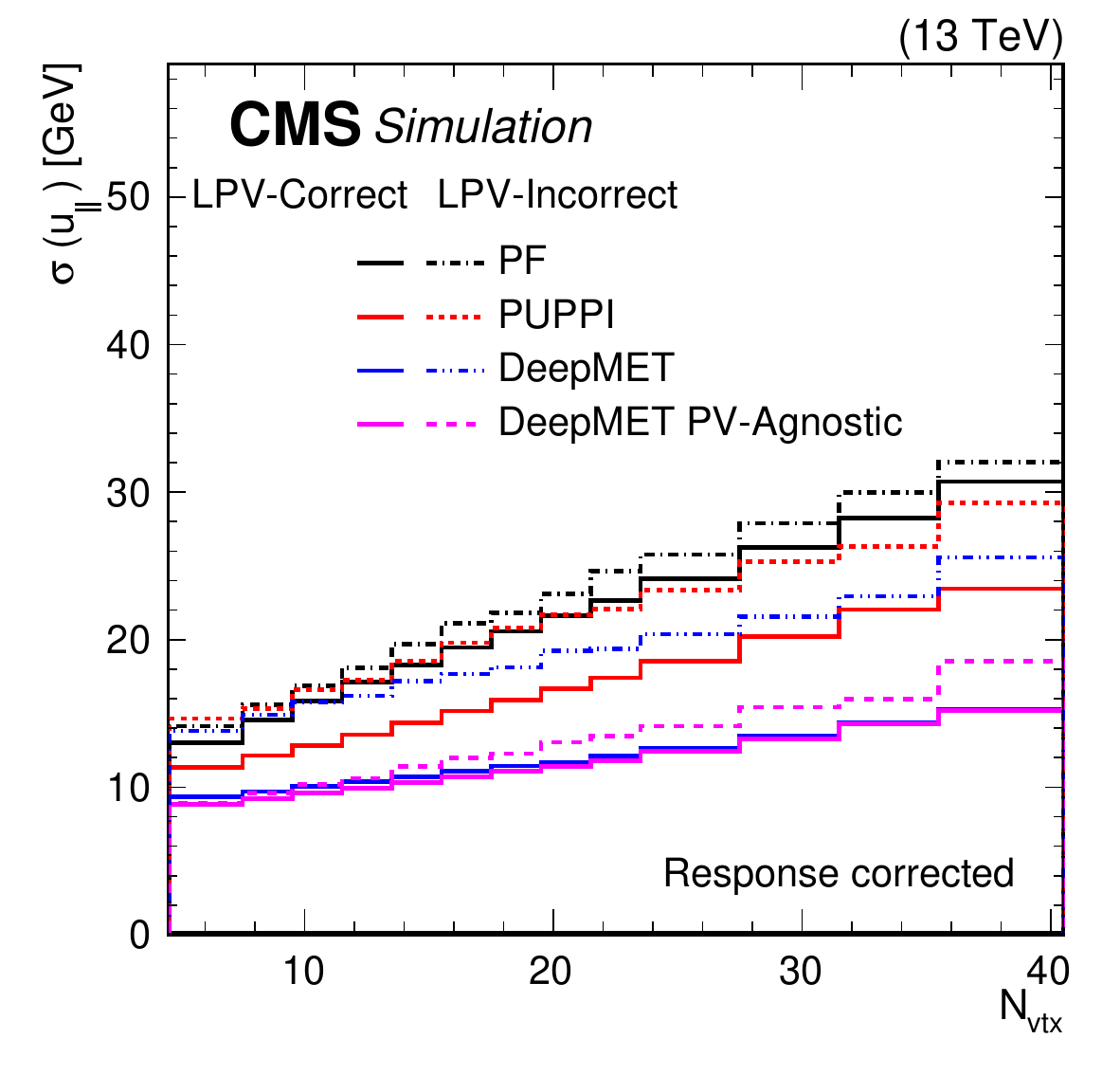}
\includegraphics[width=0.49\textwidth]{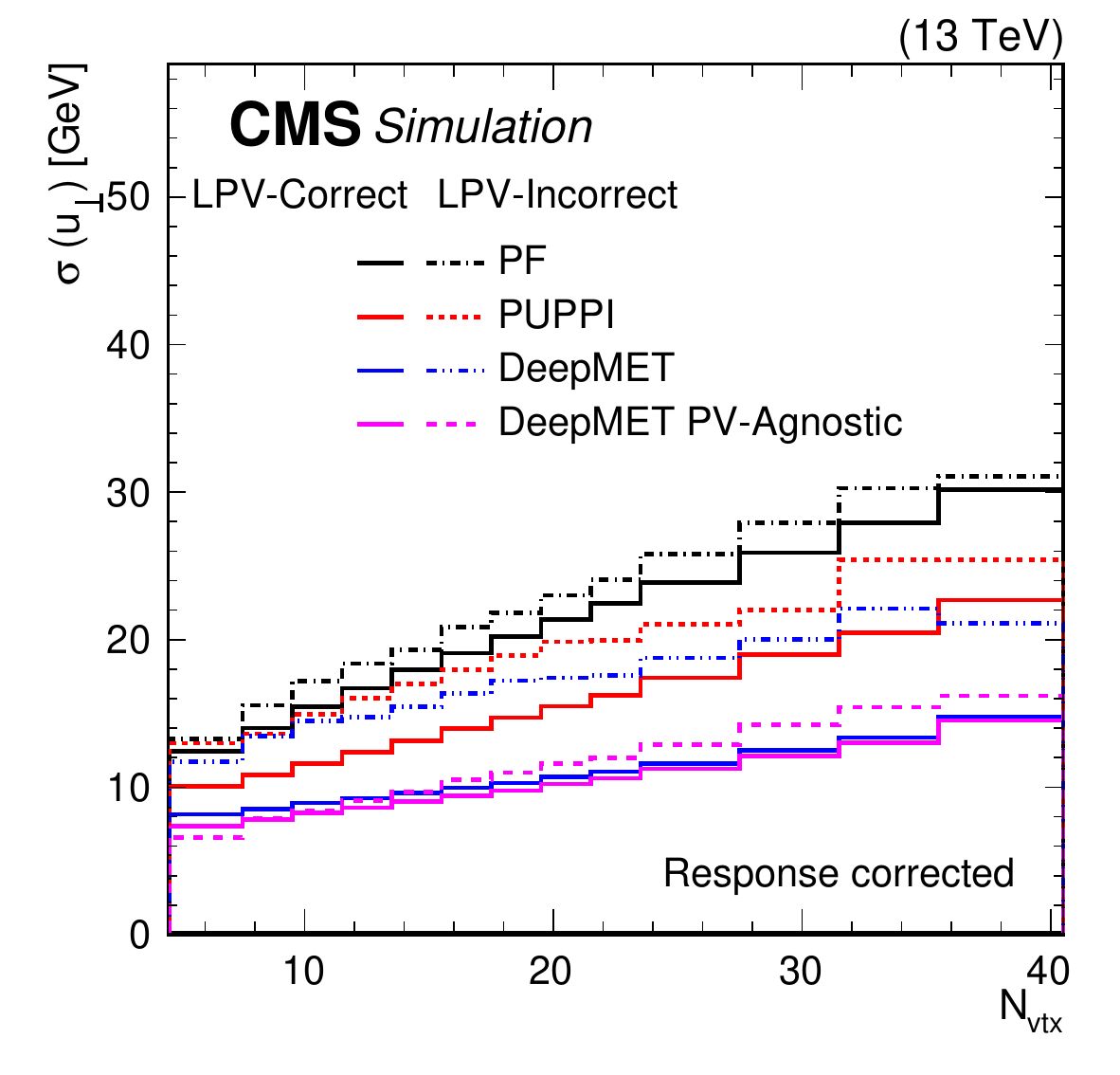}
\caption{\label{fig:wjetsmc_resol}Response (upper), response-corrected resolutions of \uparal (lower left) and \uperp (lower right) of different \ptvecmiss estimators in {\PW}+jets MC simulations. Solid (dashed) lines are from events where the LPV is (in)correctly identified.}
\end{figure*}

Figure~\ref{fig:wjetsmc_resol} illustrates the performance comparisons of various \ptvecmiss estimators for {\PW}+jets events, with solid (dashed) lines representing events where the LPV is correctly (incorrectly) identified. As anticipated, PF \ptvecmiss is the estimator least impacted by LPV identification issues. Conversely, both PUPPI \ptvecmiss and the original \DeepMET are significantly influenced. The PV-Agnostic approach helps to mitigate the dependence on LPV identifications, resulting in comparable performance between events with correct and incorrect LPV assignment. The remaining discrepancy is similar to that of PF \ptvecmiss, which is primarily due to different kinematic distributions (\eg, larger fraction of LPV-incorrect events for those with muons at higher $\abs{\eta}$).

\begin{figure}[htp]
\centering
\includegraphics[width=0.49\textwidth]{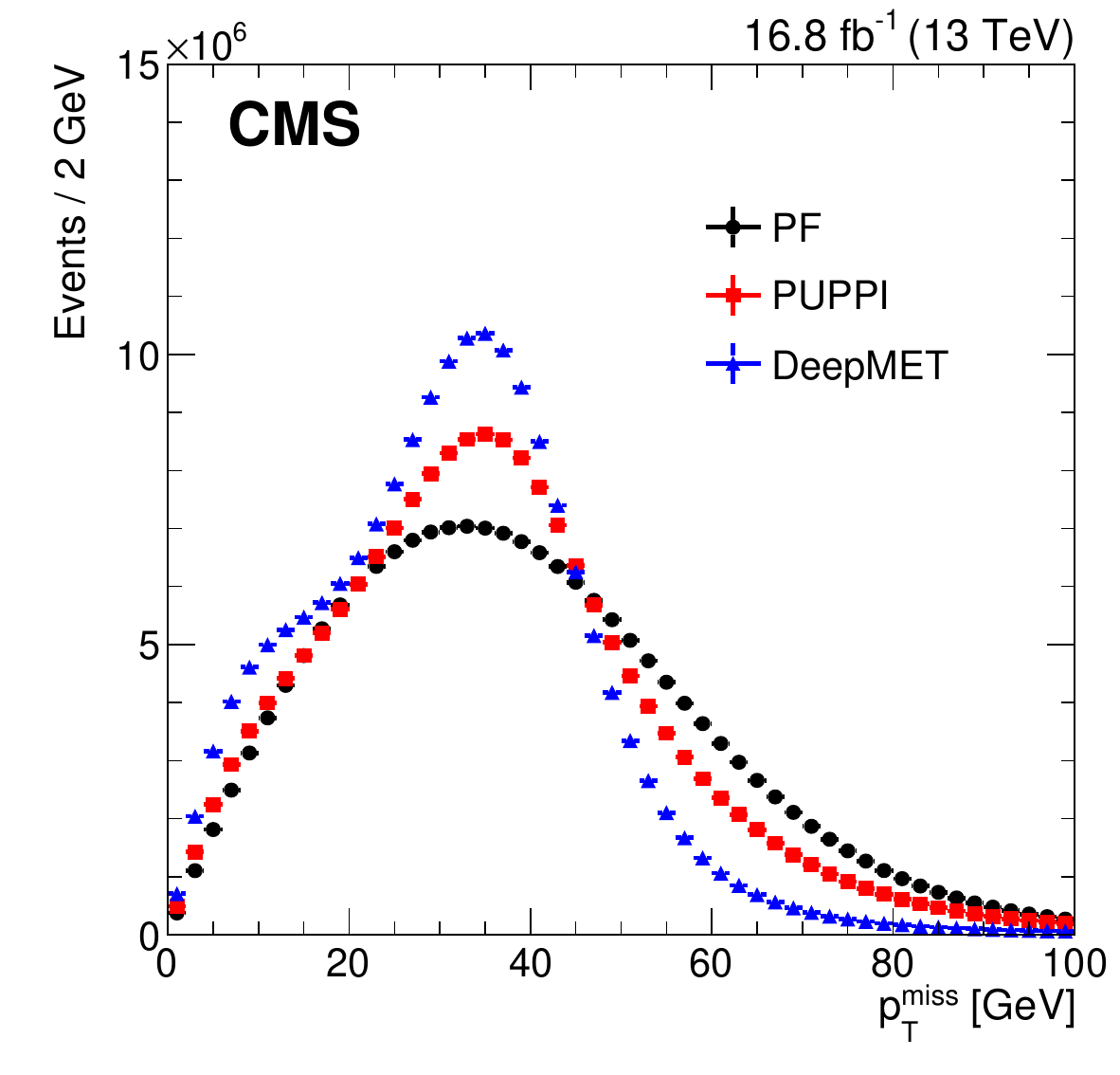}
\includegraphics[width=0.49\textwidth]{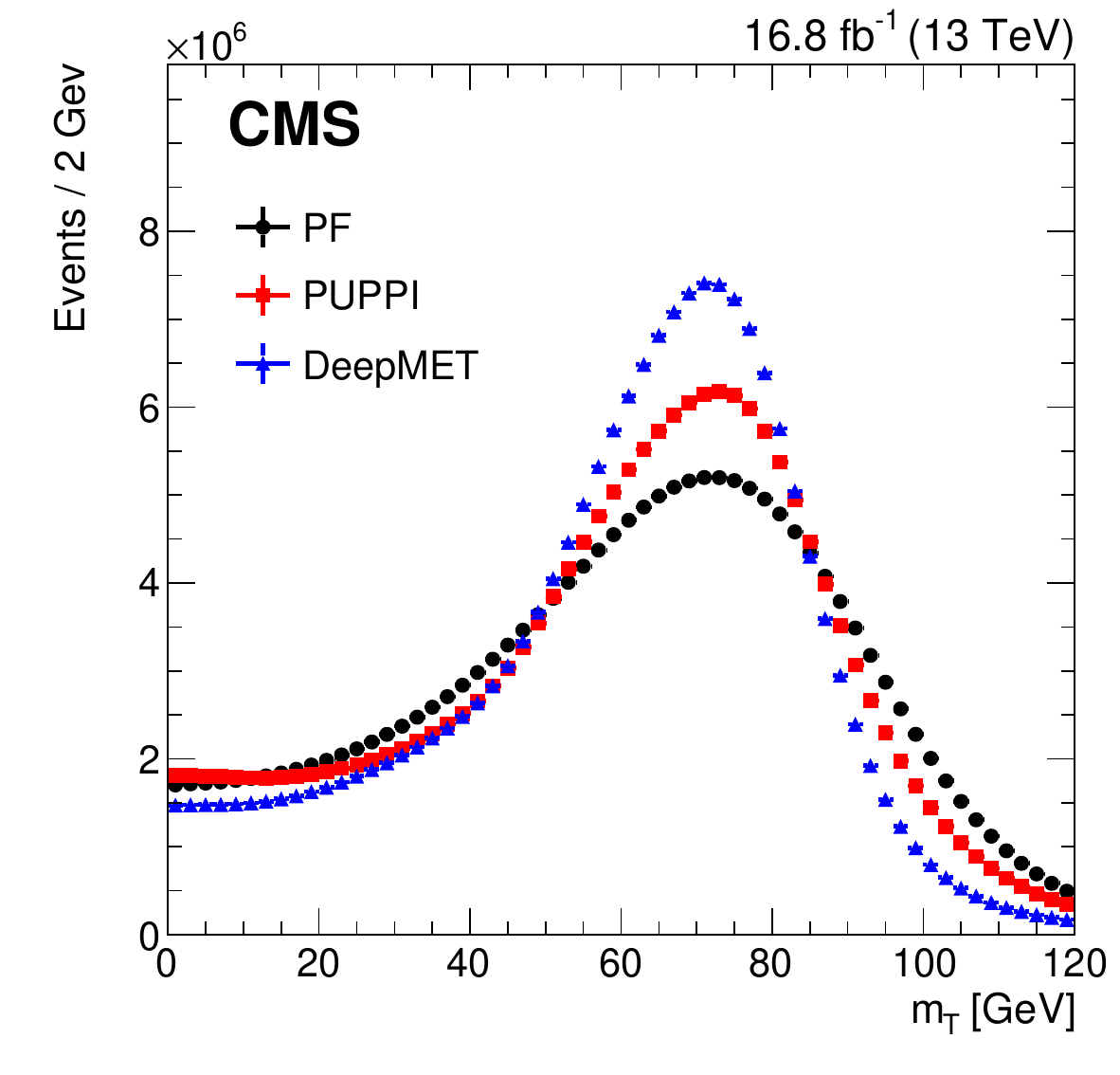}
\caption{\label{fig:wjetsmc}Distributions of \ptmiss (\cmsLeft) and $m_\mathrm{T}$ (\cmsRight) of different \ptvecmiss estimators in data after $\PW\to\mu\nu$ selections.}
\end{figure}

The \ptvecmiss and \PW boson transverse mass (defined as $m_\mathrm{T}=\sqrt{\smash[b]{2\pt^{\mu}\ptmiss[1-\cos(\Delta\phi)]}}$, where $\Delta\phi$ is the azimuthal angle difference between muon \pt and \ptvecmiss) distributions are shown in Fig.~\ref{fig:wjetsmc} for data after the $\PW\to\mu\nu$ selections. 
Table~\ref{tab:resols_wjets} gives a \ptvecmiss resolution metric: the higher half width at half maximum, defined as the difference between the value of \ptmiss ($m_\mathrm{T}$) at the \ptmiss ($m_\mathrm{T}$) probability density function (pdf) distribution maximum and the larger of the two values whose pdf is half the maximum value. For both variables, \DeepMET (PV-Agnostic) has a sharper Jacobian peak and thus provides better resolution.

\begin{table}[htb]
\topcaption{Higher half width at half maximum in \ptmiss and $m_\mathrm{T}$ in data after the $\PW\to\mu\nu$ selections.}
\centering
    \begin{scotch}{lcc}
        Higher half width at half maximum / \GeVns    & \ptmiss & $m_\mathrm{T}$ \\\hline
        PF  &  25.4 & 27.0 \\
        PUPPI & 18.4 & 16.2 \\
        \DeepMET (PV-Agnostic) & 15.2 & 13.0 \\
    \end{scotch}
    \label{tab:resols_wjets}
\end{table}

\subsection{Performance for other physics processes}\label{sec:perf_other}

\begin{figure*}
\centering
\includegraphics[width=0.45\textwidth]{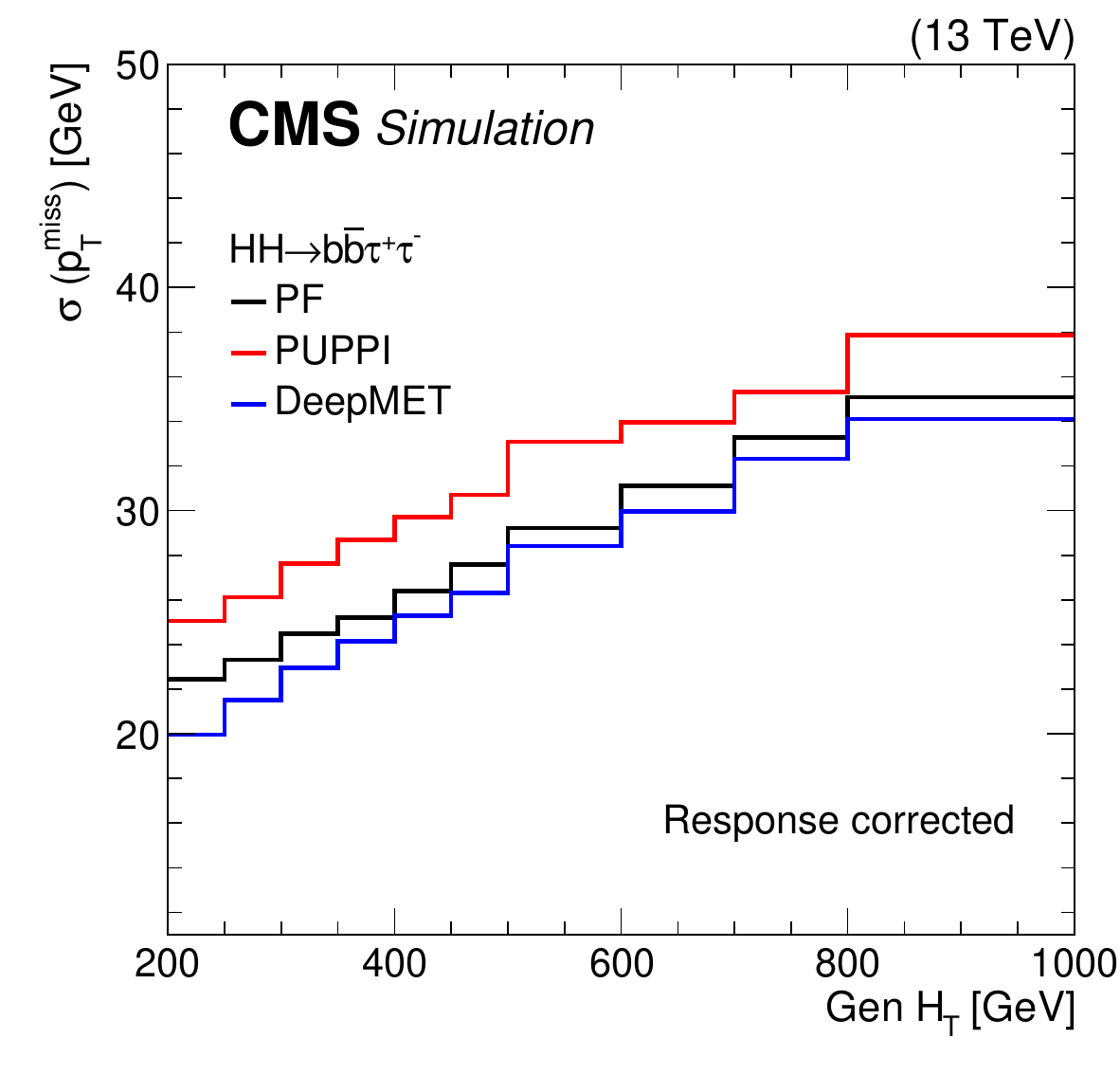}
\includegraphics[width=0.45\textwidth]{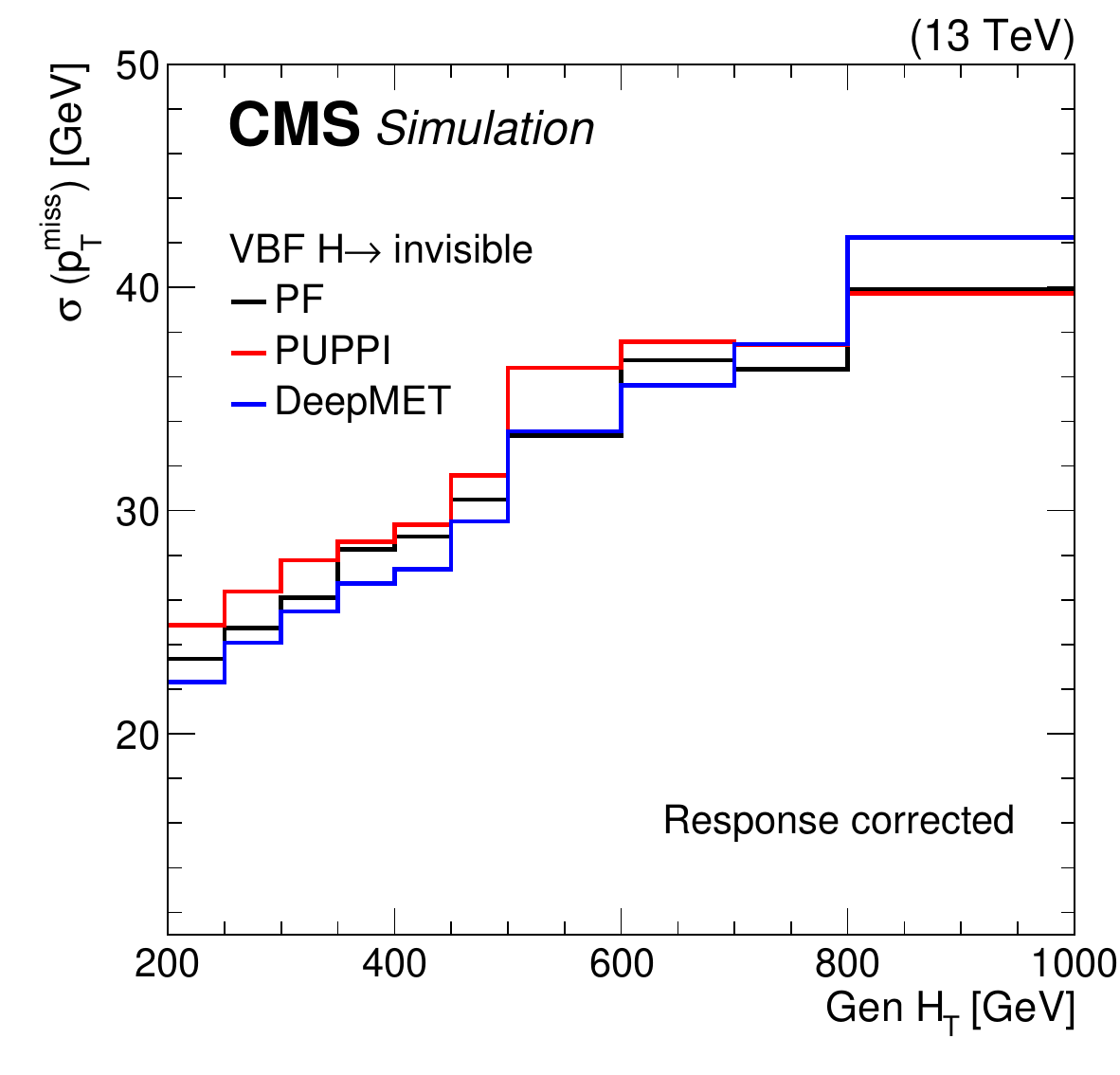}
\includegraphics[width=0.45\textwidth]{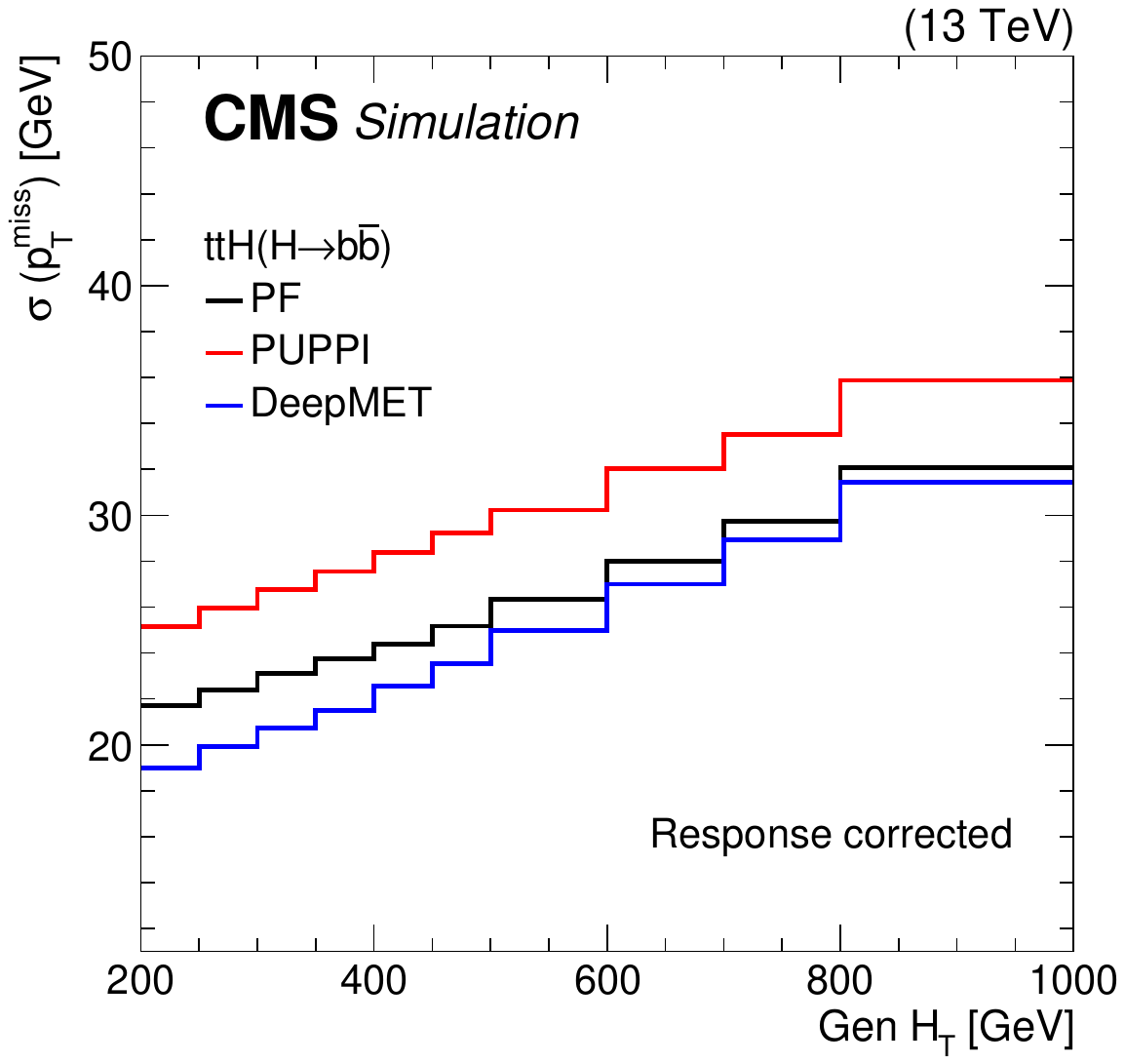}
\includegraphics[width=0.45\textwidth]{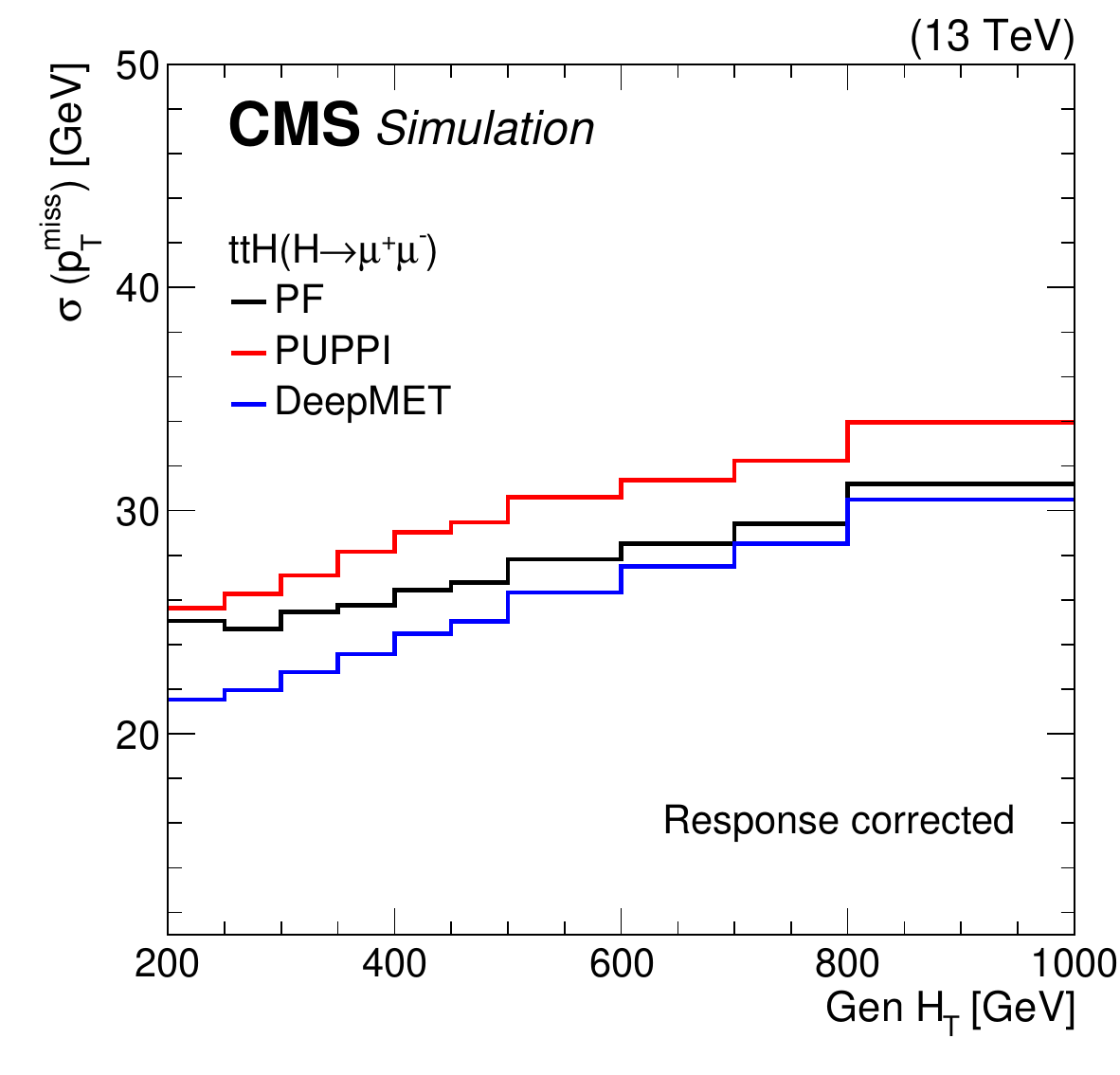}
\includegraphics[width=0.45\textwidth]{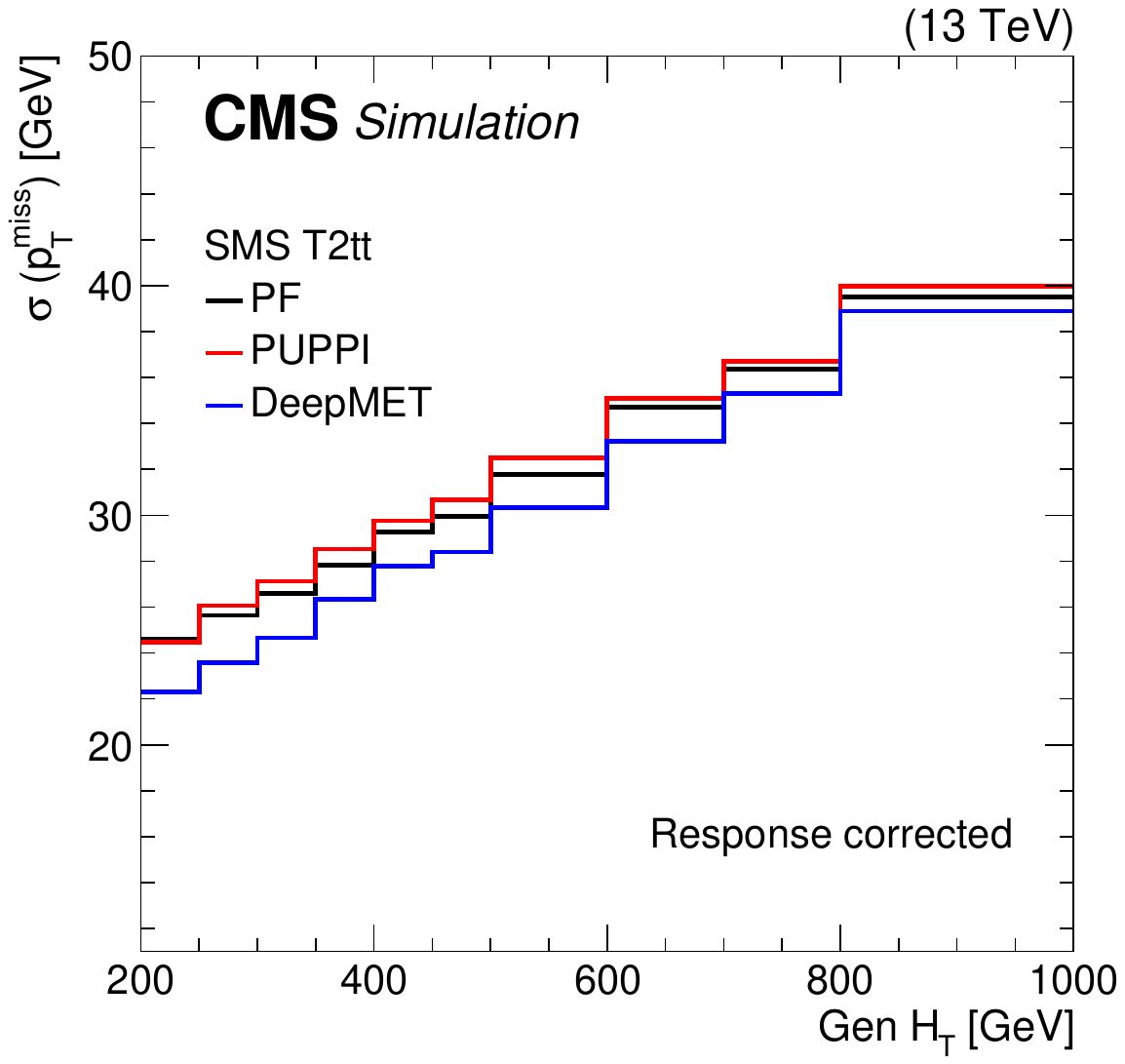}
\includegraphics[width=0.45\textwidth]{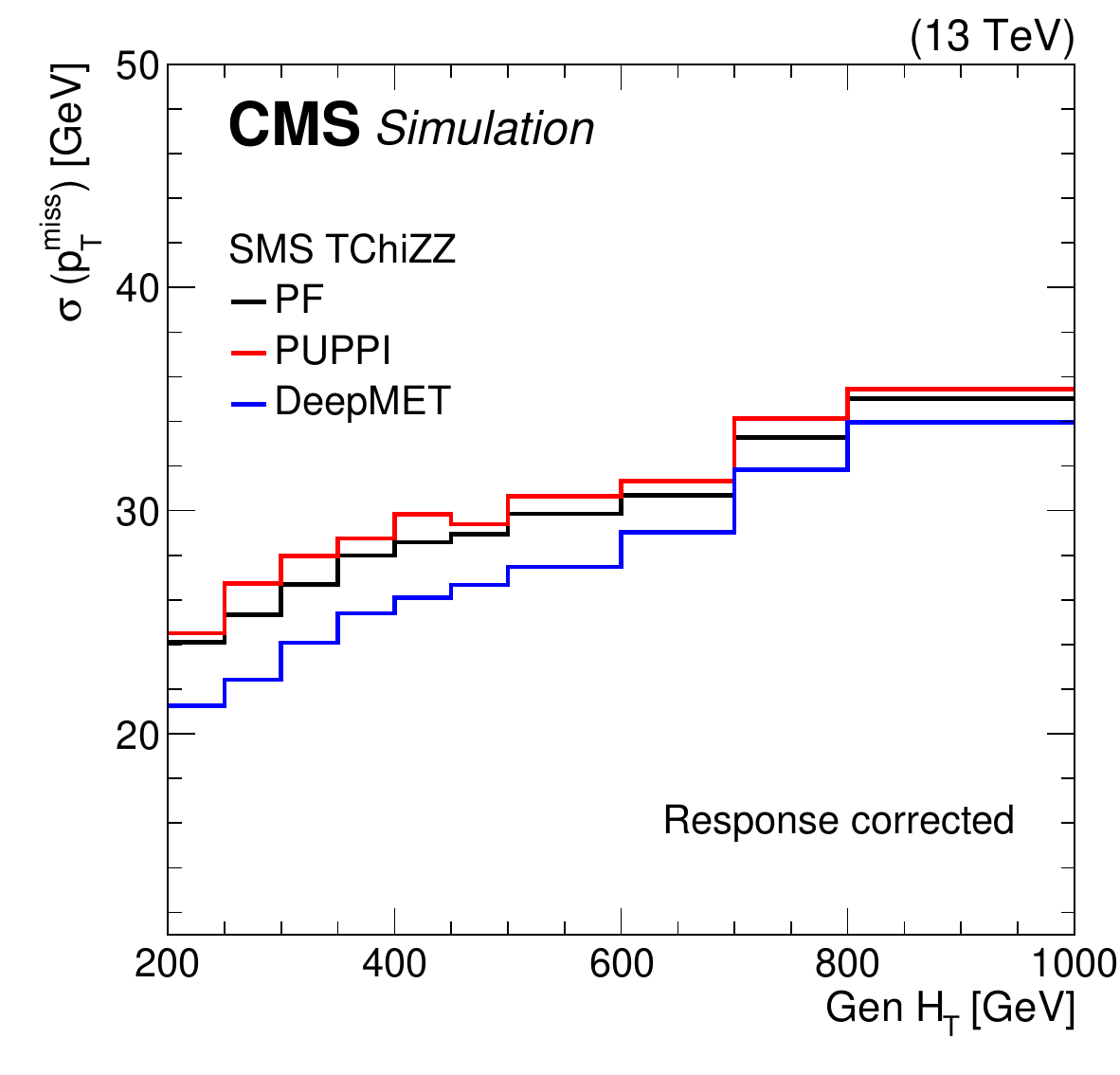}
\caption{\label{fig:deepresol_processes}Comparison of \ptmiss resolution for various physics processes in simulated events. The considered processes are $\PH\PH$ production via gluon fusion with $\PH\PH\to\PQb\PQb\PGt\PGt$ (upper left), \PH production via vector boson fusion with $\PH\to\text{invisible}$ (upper right), $\PQt\PQt\PH$ production with $\ttbar$, where both top quarks decay to leptons and either $\PH\to\PQb\PQb$ (middle left) or $\PH\to\PGm\PGm$ (middle right), the SMS T2tt process (lower left), and the SMS TChiZZ process (lower right).}
\end{figure*}

The \ptvecmiss resolution of different estimators is also compared for a number of physics processes with genuine \ptvecmiss. 
The samples are listed in Section~\ref{sec:data}, and are used without any specific event selection applied.
Figure~\ref{fig:deepresol_processes} shows the \ptmiss resolution for the \DeepMET, PUPPI \ptvecmiss, and PF \ptvecmiss algorithms as a function of the sum of the \pt's of generated jets (\HT) for various physics processes for which \ptmiss is expected to have a crucial role in the event selection and/or reconstruction, and hence for the final physics sensitivity.

\DeepMET demonstrates superior resolution in \ptmiss for all considered physics processes, including Higgs boson production with decays to SM particles, Higgs boson decay to dark matter candidates, and the supersymmetric processes described in Section~\ref{sec:data}. These processes include a wide range of different jet and lepton multiplicities, jet flavors, and cover a broad kinematic range.
The relative improvement in resolution is largest for low values of generator jet \HT, \ie, for low amounts of hard hadronic activity, indicating that \DeepMET particularly helps with the softer contributions to \ptvecmiss.

\section{\label{sec:calib}The \DeepMET calibrations}
The \DeepMET calibrations can be derived from {\PZ}+jets events by matching the performance in MC simulations to data. These corrections can be applied to the DNN outputs. 

Since the calibrations are only applied to MC simulations, all the performance results on data shown in the previous sections are unchanged after calibration.

\subsection{The XY correction}
\begin{figure}[htp]
  \centering
  \includegraphics[width=0.49\textwidth]{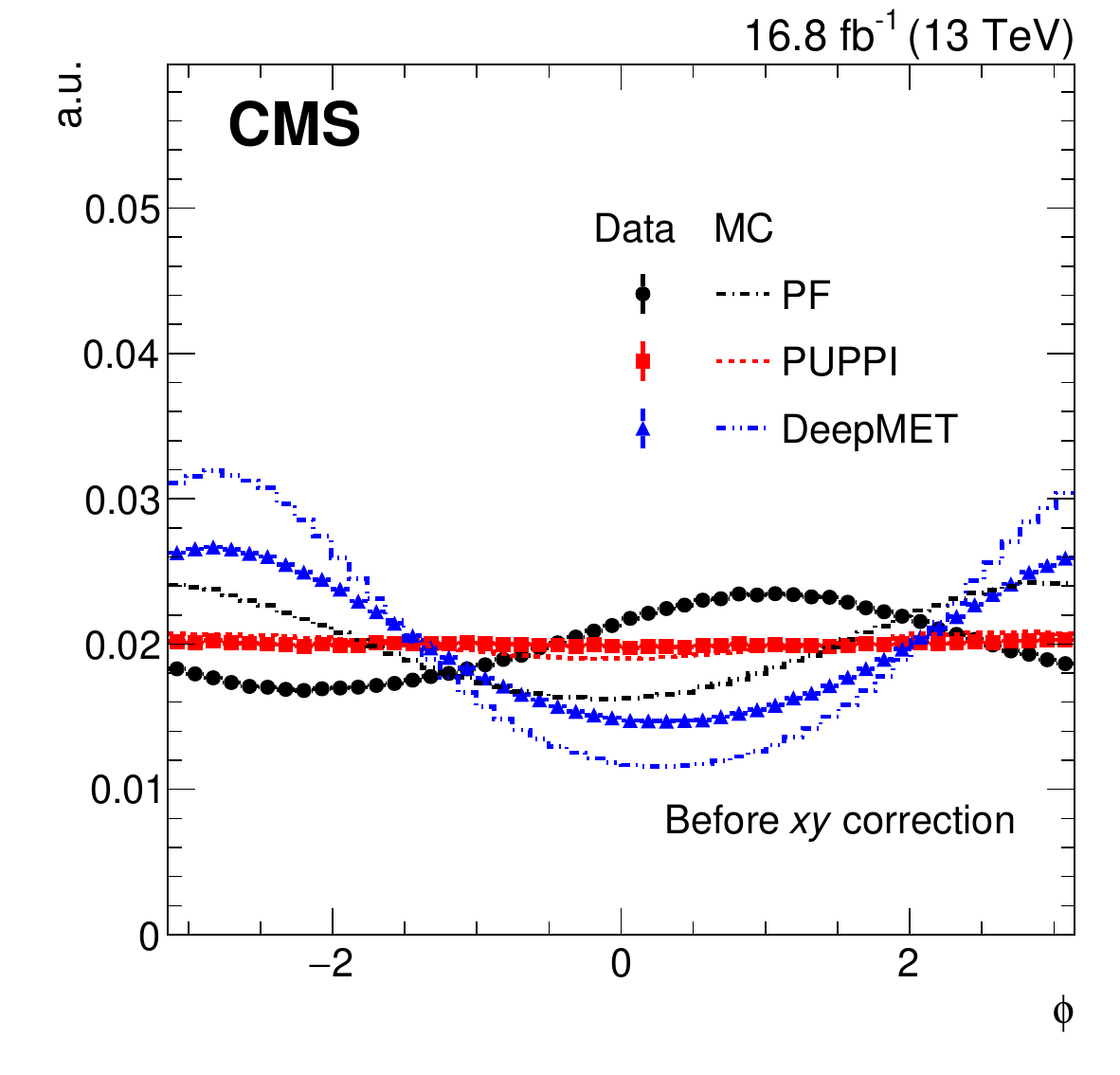}
  \includegraphics[width=0.49\textwidth]{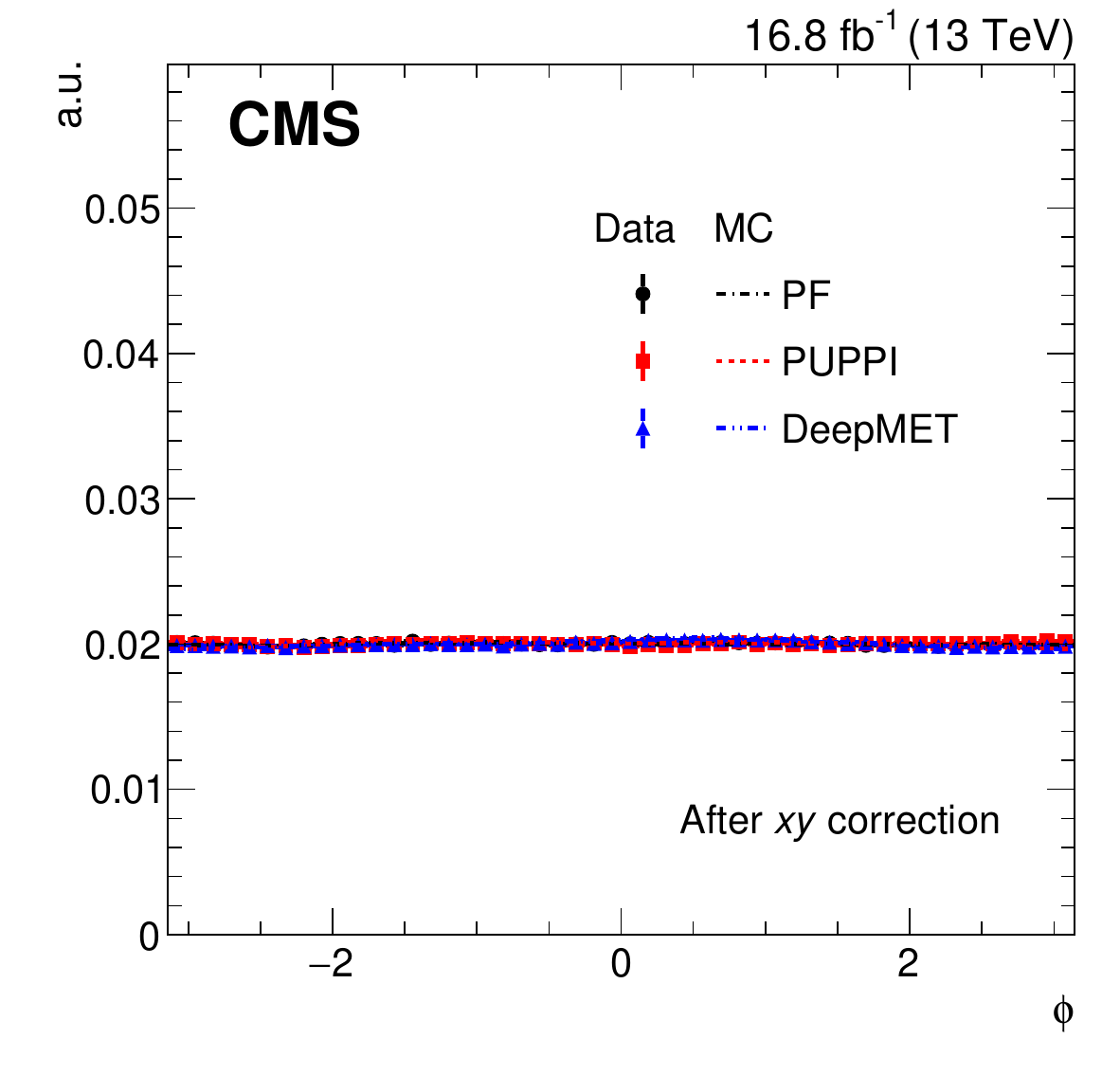}
  \caption{\label{fig:xycorr}The $\phi$ distribution of different \ptvecmiss estimators before (\cmsLeft) and after (\cmsRight) the $xy$ corrections, in data (markers) and MC simulations (dashed) after the $\PZ\to\PGm\PGm$ selections.}
\end{figure}

 Due to asymmetries in the detector responses and resolutions, and also because of the offset of the beam spot position, \etc, the reconstructed \ptvecmiss response depends on $\phi$. The average values of \ptvecmiss in the $x$ and $y$ directions are calculated as a function of the number of PVs and then subtracted from the individual \ptvecmiss components, centering the distributions at zero. The $\phi$ distributions of different \ptvecmiss estimators before and after the $xy$ correction are shown in Fig.~\ref{fig:xycorr}. The distribution becomes flat, as expected, for both data and MC simulations after correction.

\subsection{Scale and resolution calibration}
Since the jet and PF candidate resolutions are typically worse in data than predicted from simulation, a quantile-based calibration is performed.  The calibration is done in four main steps.

First the data and MC simulations are binned in \qt of the \PZ boson. For the data, the contributions from \ttbar and diboson production are estimated from MC simulation and subtracted.
Each bin's  \uparal and \uperp distribution is fitted with a double-Gaussian function. 

In the second step, each of the fitted double-Gaussian functions is integrated to get the corresponding cumulative distribution functions (CDFs). 

In MC simulation, the \uparal and \uperp CDF values in the $i$th \qt bin can be found from:
\begin{equation}
\begin{aligned}
p_{\parallel}&=\mathrm{CDF}^{\uparal, \text{MC}}_{i}(\uparal^\text{MC}),\\
p_{\perp}&=\mathrm{CDF}^{\uperp, \text{MC}}_{i}(\uperp^\text{MC}).
\end{aligned}
\end{equation}
In data, the corresponding \uparal and \uperp CDF in the same \qt bin is given by:
\begin{equation}
\begin{aligned}
\uparal^\text{data}&=(\mathrm{CDF}^{\uparal, \text{data}}_{i})^{-1}(p_{\parallel}),\\
\uperp^{\text{data}}&=(\mathrm{CDF}^{\uperp, \text{data}}_{i})^{-1}(p_{\perp}).
\end{aligned}
\end{equation}
Finally the differences $\uparal^\text{data}-\uparal^\text{MC}$ and $\uperp^\text{data}-\uperp^\text{MC}$ are applied as the corrections to \uparal and \uperp in simulation, respectively. The corrected \DeepMET is then calculated from the corrected \uparal and \uperp.

\subsection{Data \vs simulation comparisons after calibration}
\begin{figure*}[htp]
  \centering
  \includegraphics[width=0.49\textwidth]{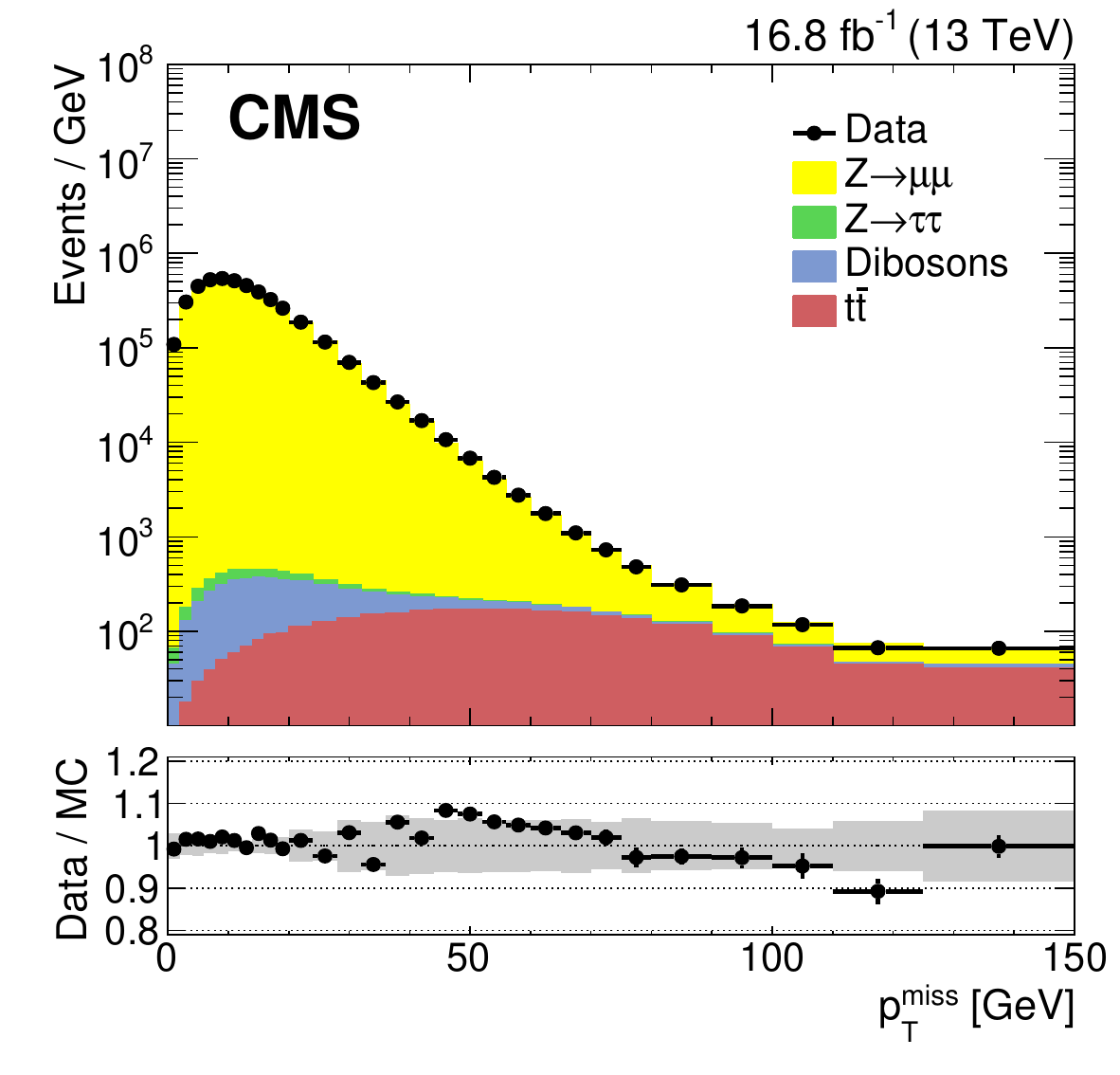}
  \includegraphics[width=0.49\textwidth]{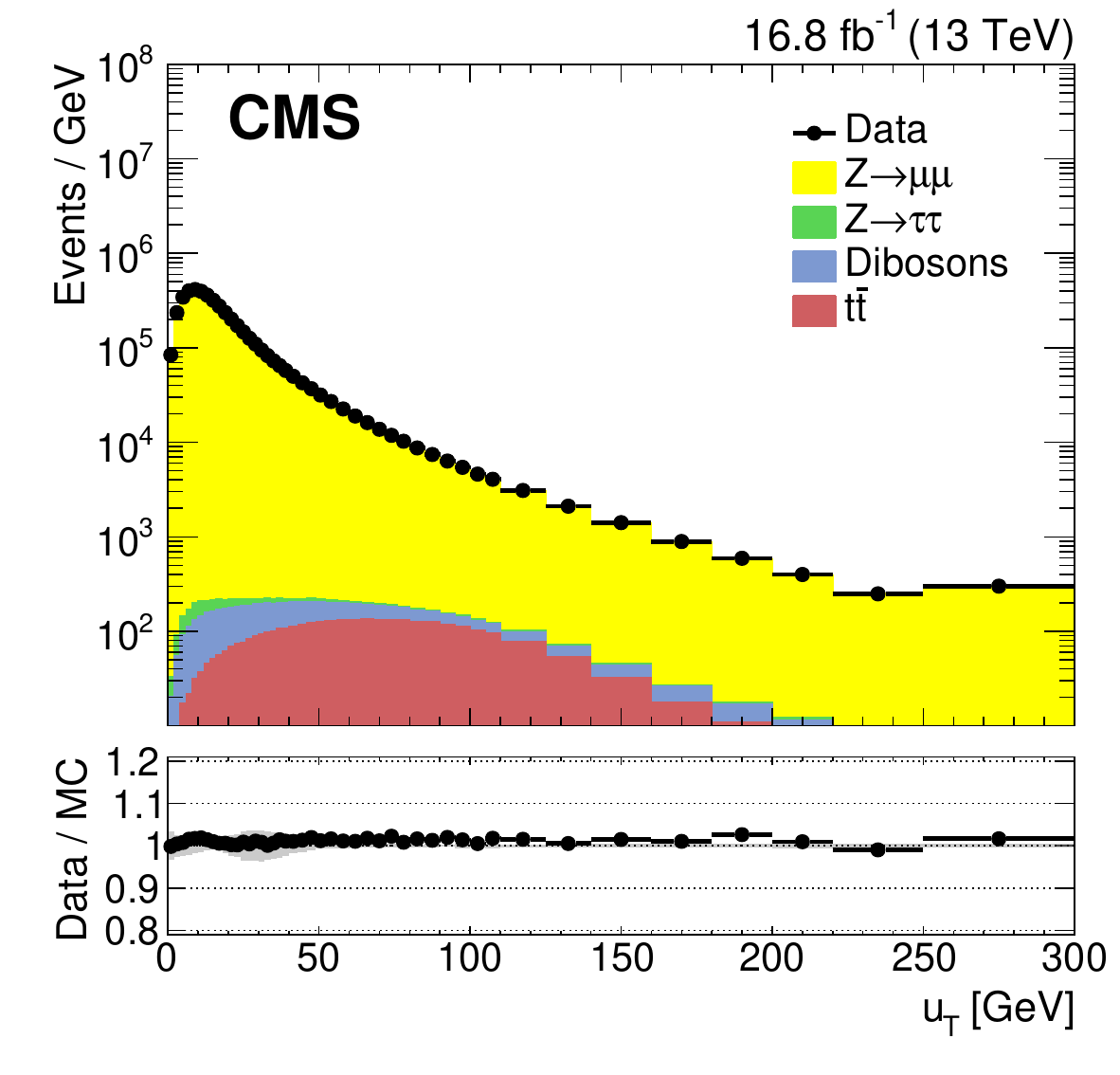}
  \includegraphics[width=0.49\textwidth]{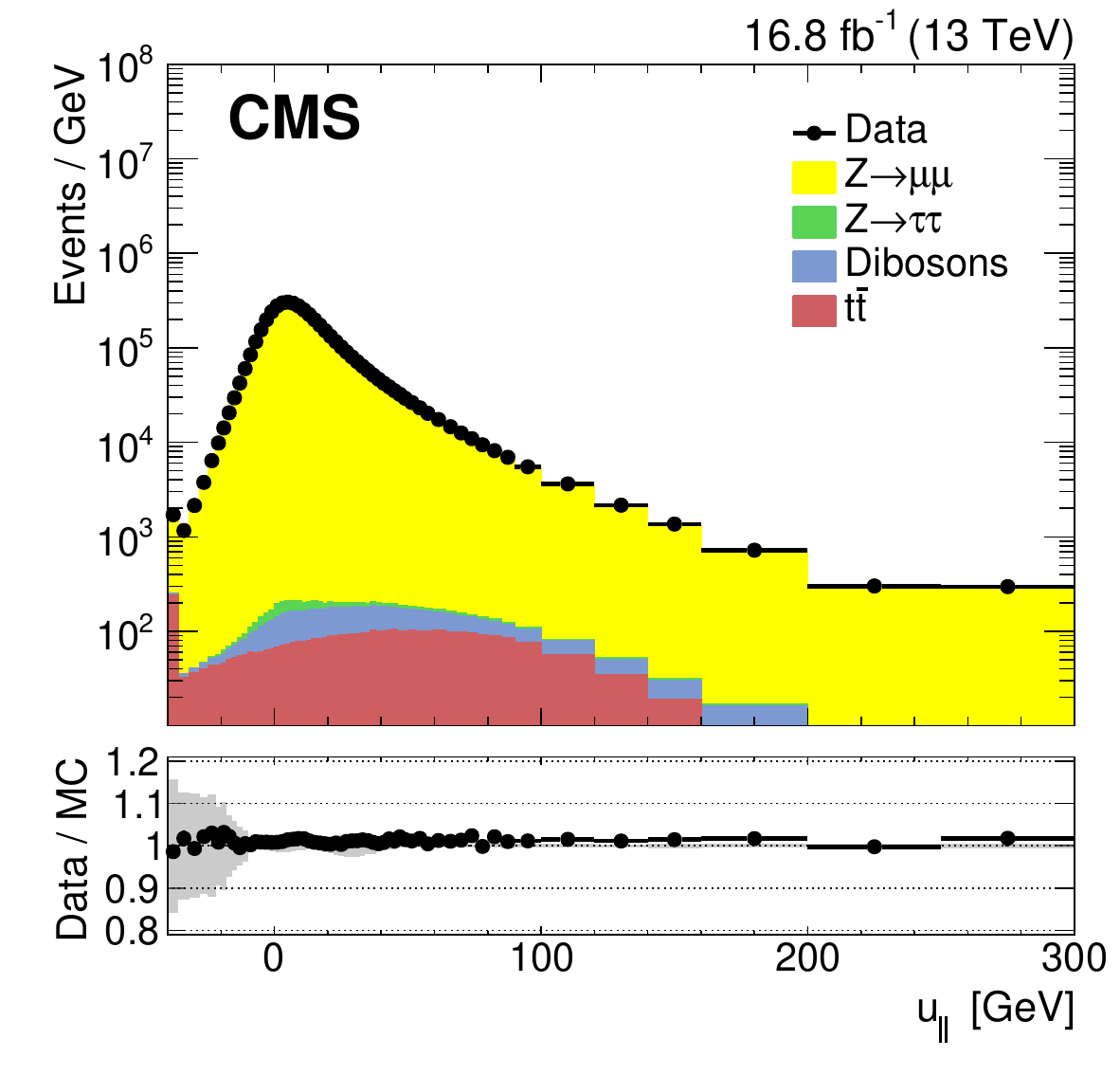}
  \includegraphics[width=0.49\textwidth]{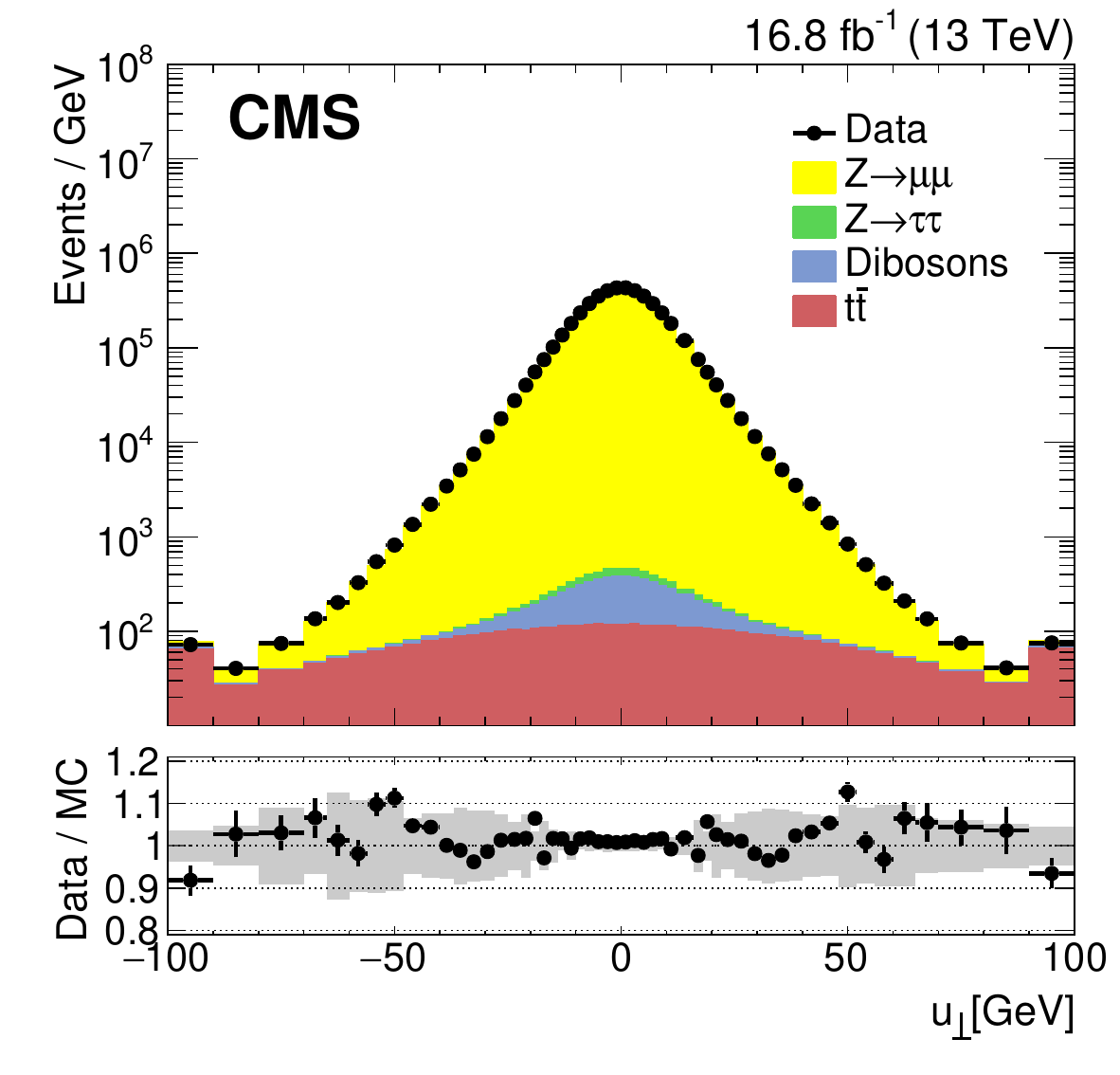}
  \caption{\label{fig:dataMCpostCorr}Data-to-simulation comparisons of \DeepMET \ptmiss (upper left), recoil \pt (upper right), \uparal (lower left), and \uperp (lower right) after the quantile correction. The underflow (overflow) contents are included in the first (last) bin. The gray band represents the systematic uncertainties discussed in Section~\ref{sec:calib:syst}.}
\end{figure*}

Figure~\ref{fig:dataMCpostCorr} illustrates the data-simulation comparisons for \DeepMET \ptmiss (upper left), recoil \pt (upper right), \uparal (lower left), and \uperp (lower right) in {\PZ}+jets events, after applying the calibrations. A \PZ boson \pt reweighting is applied to align the boson \pt distribution between data and simulation. Good agreement between data and simulation after the calibrations is observed within the systematic uncertainties, which are detailed in Section~\ref{sec:calib:syst}. 

\subsection{Systematic uncertainties from calibration}\label{sec:calib:syst}

The systematic uncertainties affecting \DeepMET are estimated during the calibration process. Three main sources are considered:
\begin{itemize}
\item \textit{Jet multiplicity binning}: in the default setup, the \PZ boson \pt is used to derive corrections in different bins. One alternative choice is to have bins in both the \pt of the \PZ boson and CHS jet multiplicity. A new set of corrections are derived using the same setup but with two-dimensional binning, and the difference between using both the \pt of the \PZ boson and the jet multiplicity, and using only the \pt of the \PZ boson, is chosen as one systematic uncertainty.
\item \textit{Choice of the modeling function}: in the default setup, a double-Gaussian function is chosen to model the \uparal and \uperp distributions. To estimate the possible bias due to this choice, a Gaussian smoothing function is used to smooth the \uparal and \uperp distributions, and then the CDFs and corrections are derived using the smoothed distributions. The correction difference between the double-Gaussian and Gaussian smoothing functions is chosen as one systematic uncertainty. 
\item \textit{Background estimation and subtraction}. When fitting the data distributions, the background contributions from \ttbar and diboson processes are estimated using MC simulation and then subtracted from data. To estimate the systematic uncertainties from the background subtraction, the contributions estimated from background MC samples are varied by $\pm10\%$ (typical cross section measurement and theoretical uncertainties), and the difference between the calibrations using the central values and using these are taken as the systematic uncertainty.
\end{itemize}

The total uncertainties are shown in the gray band of Fig.~\ref{fig:dataMCpostCorr}.
Around the peak of the {\PZ}+jets distributions (low \ptmiss and smaller $\abs{\uperp}$), the relative difference is around a few percent. In the tail regions the uncertainty can be as large as 10--15\%. Similar studies have also been carried out in the $\ttbar$ region by selecting two muons and one b jet, where good data-to-simulation agreement has also been observed after applying the corrections.

\section{\label{sec:deeper}Training without PUPPI weights}

To understand the role and the importance of the PUPPI weights in \DeepMET's performance, a version of \DeepMET is trained without the PUPPI weight information. Because of the DNN nature of the architecture, particle-level information is not shared between different PF candidates, and the PUPPI weight is the \textit{only} input that contains local neighborhood information. This comparison sheds light on the relative importance of a particle's neighborhood in the observed performance improvement.

\begin{figure*}[htp]
   \centering
    \includegraphics[width=0.49\textwidth]{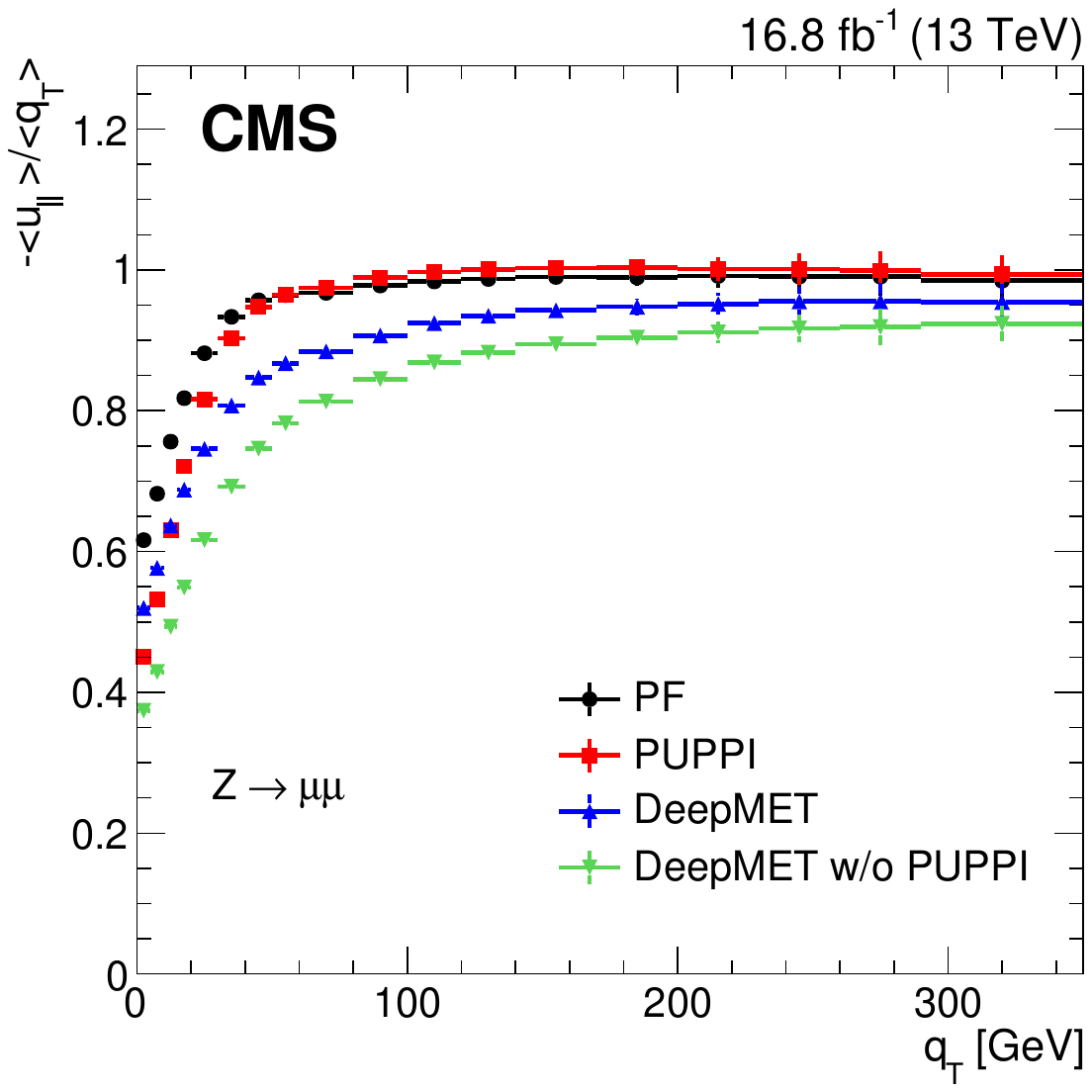}\\
    \includegraphics[width=0.49\textwidth]{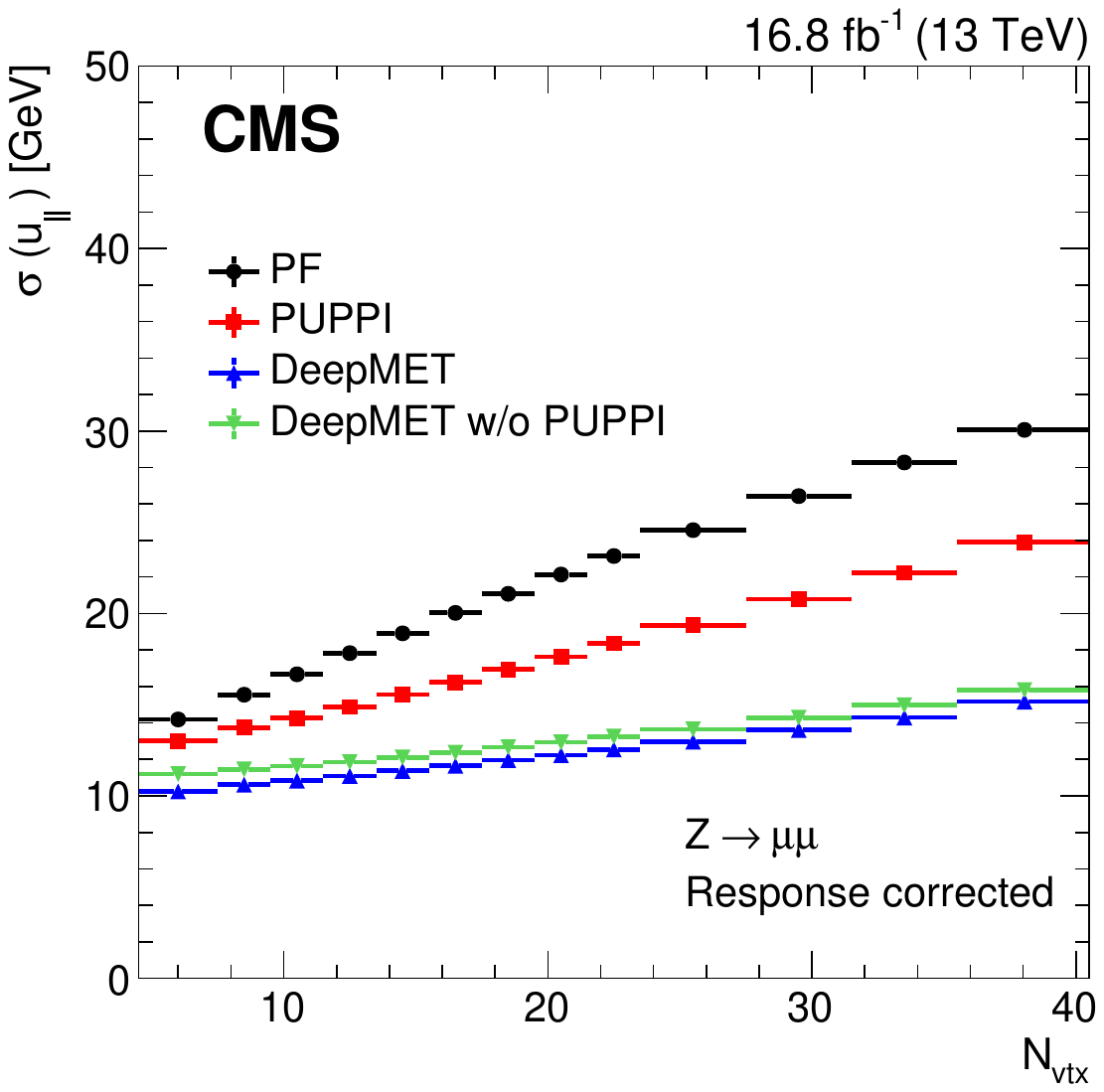}
    \includegraphics[width=0.49\textwidth]{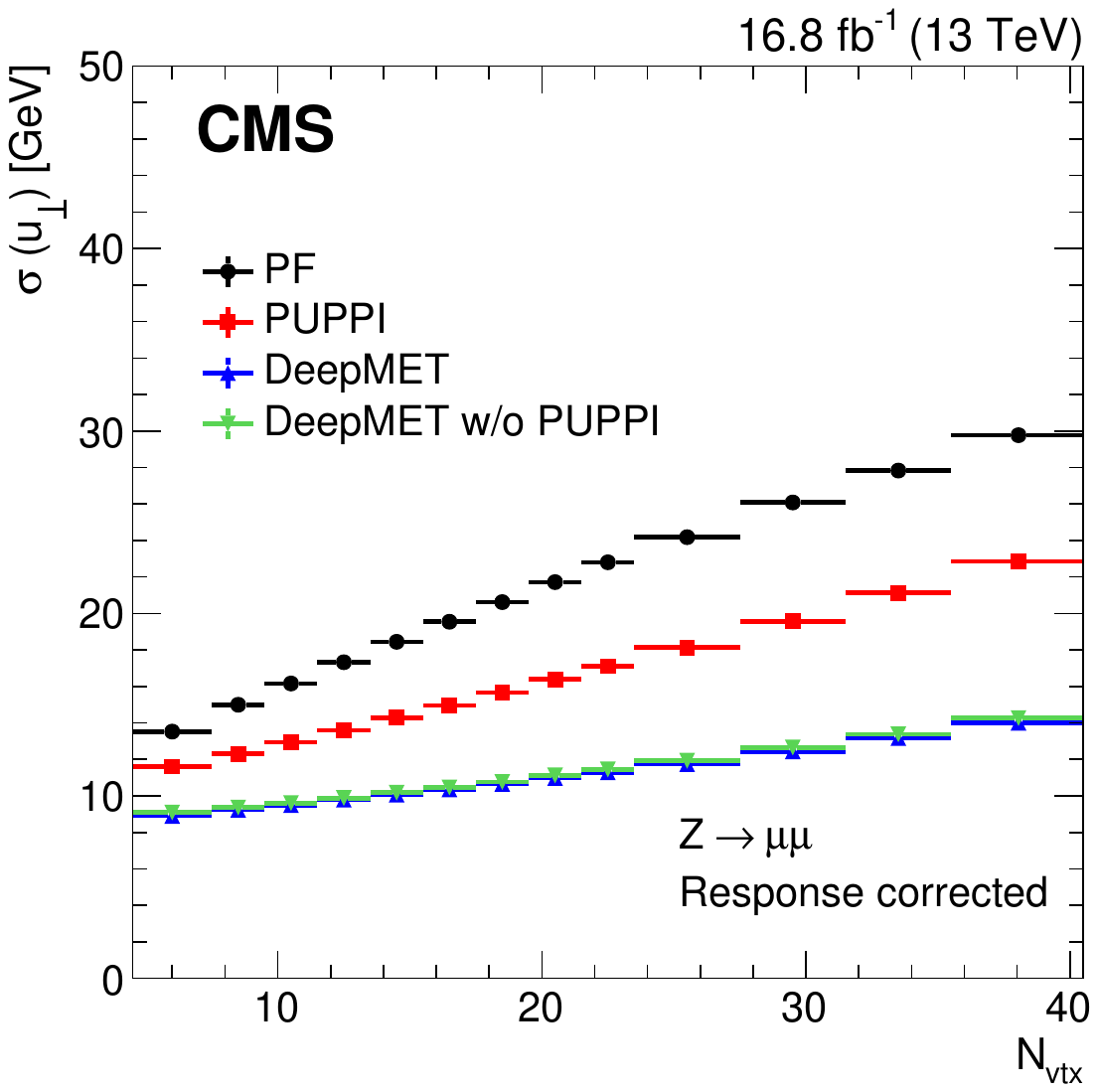}
    \caption{\label{fig:deepresol_nopuppi}Response (upper) and response-corrected resolutions of \uparal (lower left) and \uperp (lower right) of different \ptvecmiss estimators in data after the $\PZ\to\PGm\PGm$ selections.}
\end{figure*}

Figure~\ref{fig:deepresol_nopuppi} shows comparisons of the response and response-corrected resolution of \DeepMET trained with (blue) and without (green) PUPPI weights, along with PF \ptvecmiss and PUPPI \ptvecmiss, after $\PZ\to\PGm\PGm$ selections. As can be seen for the $\PZ\to\PGm\PGm$ physics process, PUPPI weights mostly contribute to response improvements; the resolutions with and without PUPPI weights are similar.

\section{\label{sec:implementation}Computational performance}
The model trained using \KERAS is stored as a frozen graph in \TENSORFLOW format and integrated into the CMS software framework \CMSSW for subsequent inference processes. The \CMSSW framework is an open-source software framework utilized in triggering, data reconstruction, simulation, and offline analysis~\cite{CMS:2006myw,cmssw}. In this setup, the number of input PF candidates is set to 4500, determined as the maximum count of PF candidates under the current operational conditions (typically ranging from 1500 to 2000). The model inputs consist of the 11 features of each PF candidate, as discussed in Section~\ref{sec:algo}. In cases where there are fewer PF candidates, the input features are padded with zeros. The model processes these inputs to compute \ptvecmiss in the $x$ and $y$ directions as the outputs. \DeepMET inference can be performed at any stage of data processing, as long as the list of PF candidates is provided.

The computational time is also examined. When running with one thread on a single CPU core, the average processing time for the \DeepMET inference within \CMSSW is approximately 10\unit{ms}~\cite{CMS:2024twn}, while the total event processing time ranges from 1 to 10\unit{s}, depending on the event complexity and the associated data tier. Due to the compact size of the model, the average processing time for \DeepMET is much shorter than the overall \CMSSW processing time.

\section{\label{sec:summary}Summary}
This paper presents a new \ptvecmiss estimator, \DeepMET, which is based on a deep neural network. The \DeepMET algorithm utilizes each individual particle reconstructed by the CMS particle-flow algorithm as input and assigns a weight $w_i$ and two bias terms, $b_{i,x}$ and $b_{i,y}$, to each candidate. The estimated \ptvecmiss is the negative vector sum of the weighted transverse momenta of all candidates, plus their bias contributions. 

With 4541 trainable parameters, the training and deployment of \DeepMET is computationally efficient.
\DeepMET is trained using {\PZ}+jets and \ttbar samples, but achieves 10--30\% better resolution compared with the current PF and PUPPI \ptvecmiss estimators across multiple physics processes, such as {\PZ}+jets, {\PW}+jets, Higgs boson production, and processes with dark matter candidates.
Another important feature of the \DeepMET algorithm is its high resilience to pileup, improving the physics reach in LHC Run 2 and Run 3, and future High-Luminosity LHC conditions.
Specifically for the measurement of the \PW boson mass, a PV-Agnostic version of \DeepMET is designed to be more robust in $\PW \to \PGm\PGn$ events where the LPV is not always identified correctly. Events containing $\PZ \to \PGm\PGm$ decays are used to calibrate \ptvecmiss, including corrections for asymmetries in detector response, as well as for \ptvecmiss scale and resolution. 

Good agreement between data and simulation is found.
The \DeepMET estimator demonstrates the potential to improve the precision of SM measurements and to achieve higher sensitivity in beyond the standard model physics searches.

\begin{acknowledgments}
We congratulate our colleagues in the CERN accelerator departments for the excellent performance of the LHC and thank the technical and administrative staffs at CERN and at other CMS institutes for their contributions to the success of the CMS effort. In addition, we gratefully acknowledge the computing centers and personnel of the Worldwide LHC Computing Grid and other centers for delivering so effectively the computing infrastructure essential to our analyses. Finally, we acknowledge the enduring support for the construction and operation of the LHC, the CMS detector, and the supporting computing infrastructure provided by the following funding agencies: SC (Armenia), BMBWF and FWF (Austria); FNRS and FWO (Belgium); CNPq, CAPES, FAPERJ, FAPERGS, and FAPESP (Brazil); MES and BNSF (Bulgaria); CERN; CAS, MoST, and NSFC (China); MINCIENCIAS (Colombia); MSES and CSF (Croatia); RIF (Cyprus); SENESCYT (Ecuador); ERC PRG, TARISTU24-TK10 and MoER TK202 (Estonia); Academy of Finland, MEC, and HIP (Finland); CEA and CNRS/IN2P3 (France); SRNSF (Georgia); BMFTR, DFG, and HGF (Germany); GSRI (Greece); NKFIH (Hungary); DAE and DST (India); IPM (Iran); SFI (Ireland); INFN (Italy); MSIT and NRF (Republic of Korea); MES (Latvia); LMTLT (Lithuania); MOE and UM (Malaysia); BUAP, CINVESTAV, CONACYT, LNS, SEP, and UASLP-FAI (Mexico); MOS (Montenegro); MBIE (New Zealand); PAEC (Pakistan); MES, NSC, and NAWA (Poland); FCT (Portugal); MESTD (Serbia); MICIU/AEI and PCTI (Spain); MOSTR (Sri Lanka); Swiss Funding Agencies (Switzerland); MST (Taipei); MHESI and NSTDA (Thailand); TUBITAK and TENMAK (T\"{u}rkiye); NASU (Ukraine); STFC (United Kingdom); DOE and NSF (USA).

\hyphenation{Rachada-pisek} Individuals have received support from the Marie-Curie program and the European Research Council and Horizon 2020 Grant, contract Nos.\ 675440, 724704, 752730, 758316, 765710, 824093, 101115353, 101002207, 101001205, and COST Action CA16108 (European Union); the Leventis Foundation; the Alfred P.\ Sloan Foundation; the Alexander von Humboldt Foundation; the Science Committee, project no. 22rl-037 (Armenia); the Fonds pour la Formation \`a la Recherche dans l'Industrie et dans l'Agriculture (FRIA-Belgium); the Beijing Municipal Science \& Technology Commission, No. Z191100007219010, the Fundamental Research Funds for the Central Universities, the Ministry of Science and Technology of China under Grant No. 2023YFA1605804, and the Natural Science Foundation of China under Grant No. 12061141002 (China); the Ministry of Education, Youth and Sports (MEYS) of the Czech Republic; the Shota Rustaveli National Science Foundation, grant FR-22-985 (Georgia); the Deutsche Forschungsgemeinschaft (DFG), among others, under Germany's Excellence Strategy -- EXC 2121 ``Quantum Universe" -- 390833306, and under project number 400140256 - GRK2497; the Hellenic Foundation for Research and Innovation (HFRI), Project Number 2288 (Greece); the Hungarian Academy of Sciences, the New National Excellence Program - \'UNKP, the NKFIH research grants K 131991, K 133046, K 138136, K 143460, K 143477, K 146913, K 146914, K 147048, 2020-2.2.1-ED-2021-00181, TKP2021-NKTA-64, and 2021-4.1.2-NEMZ\_KI-2024-00036 (Hungary); the Council of Science and Industrial Research, India; ICSC -- National Research Center for High Performance Computing, Big Data and Quantum Computing, FAIR -- Future Artificial Intelligence Research, and CUP I53D23001070006 (Mission 4 Component 1), funded by the NextGenerationEU program (Italy); the Latvian Council of Science; the Ministry of Education and Science, project no. 2022/WK/14, and the National Science Center, contracts Opus 2021/41/B/ST2/01369, 2021/43/B/ST2/01552, 2023/49/B/ST2/03273, and the NAWA contract BPN/PPO/2021/1/00011 (Poland); the Funda\c{c}\~ao para a Ci\^encia e a Tecnologia, grant CEECIND/01334/2018 (Portugal); the National Priorities Research Program by Qatar National Research Fund; MICIU/AEI/10.13039/501100011033, ERDF/EU, "European Union NextGenerationEU/PRTR", and Programa Severo Ochoa del Principado de Asturias (Spain); the Chulalongkorn Academic into Its 2nd Century Project Advancement Project, and the National Science, Research and Innovation Fund via the Program Management Unit for Human Resources \& Institutional Development, Research and Innovation, grant B39G680009 (Thailand); the Kavli Foundation; the Nvidia Corporation; the SuperMicro Corporation; the Welch Foundation, contract C-1845; and the Weston Havens Foundation (USA).
\end{acknowledgments}

\bibliography{auto_generated}

\providecommand{\href}[2]{#2}\begingroup\raggedright\begin{thebibliography}{10}%
\makeatletter
\providecommand{\hrefCMSnoop }[0]{\@secondoftwo}%
\makeatother
\providecommand{\doi}{\texttt{doi:}\begingroup \urlstyle{tt}\Url}

\bibitem{Chatrchyan:2008zzk}
\hrefCMSnoop {}{{CMS Collaboration}, ``The {CMS} experiment at the {CERN}
  {LHC}'',} \textit{ JINST} \textbf{ 3} (2008) S08004,
  \href{http://dx.doi.org/10.1088/1748-0221/3/08/S08004}{\doi{10.1088/1748-0221/3/08/S08004}}.

\bibitem{ATLAS:2008xda}
\hrefCMSnoop {}{{ATLAS Collaboration}, ``The {ATLAS} experiment at the {CERN}
  {Large Hadron Collider}'',} \textit{ JINST} \textbf{ 3} (2008) S08003,
  \href{http://dx.doi.org/10.1088/1748-0221/3/08/S08003}{\doi{10.1088/1748-0221/3/08/S08003}}.

\bibitem{CMS:2020ebo}
\hrefCMSnoop {}{{CMS Collaboration}, ``{Pileup mitigation at CMS in 13\TeV
  data}'',} \textit{ JINST} \textbf{ 15} (2020) P09018,
  \href{http://dx.doi.org/10.1088/1748-0221/15/09/P09018}{\doi{10.1088/1748-0221/15/09/P09018}},
  \href{http://www.arXiv.org/abs/2003.00503}{\texttt{arXiv:2003.00503}}.

\bibitem{Bertolini:2014bba}
\hrefCMSnoop {}{D.~Bertolini, P.~Harris, M.~Low, and N.~Tran, ``Pileup per
  particle identification'',} \textit{ JHEP} \textbf{ 10} (2014) 059,
  \href{http://dx.doi.org/10.1007/JHEP10(2014)059}{\doi{10.1007/JHEP10(2014)059}},
\href{http://www.arXiv.org/abs/1407.6013}{\texttt{arXiv:1407.6013}}.

\bibitem{CMS-DP-2021-001}
\href {https://cds.cern.ch/record/2751563}{{CMS Collaboration},
  ``{Pileup-per-particle identification: optimisation for Run 2 Legacy and
  beyond}'',} CMS Detector Performance Summary CMS-DP-2021-001, 2021.

\bibitem{CMS:2014xym}
\hrefCMSnoop {}{{CMS Collaboration}, ``{Performance of the CMS missing
  transverse momentum reconstruction in pp data at $\sqrt{s}$ = 8 TeV}'',}
  \textit{ JINST} \textbf{ 10} (2015) P02006,
  \href{http://dx.doi.org/10.1088/1748-0221/10/02/P02006}{\doi{10.1088/1748-0221/10/02/P02006}},
  \href{http://www.arXiv.org/abs/1411.0511}{\texttt{arXiv:1411.0511}}.

\bibitem{CMS:2014wdm}
\hrefCMSnoop {}{{CMS Collaboration}, ``{Evidence for the 125 GeV Higgs boson
  decaying to a pair of $\tau$ leptons}'',} \textit{ JHEP} \textbf{ 05} (2014)
  104,
  \href{http://dx.doi.org/10.1007/JHEP05(2014)104}{\doi{10.1007/JHEP05(2014)104}},
  \href{http://www.arXiv.org/abs/1401.5041}{\texttt{arXiv:1401.5041}}.

\bibitem{hepdata}
\hrefCMSnoop {}{``{HEPD}ata record for this analysis'',} 2025.
\newblock
  \href{http://dx.doi.org/10.17182/hepdata.159179}{\doi{10.17182/hepdata.159179}}.

\bibitem{CMS:2020cmk}
\hrefCMSnoop {}{{CMS Collaboration}, ``{Performance of the CMS Level-1 trigger
  in proton-proton collisions at $\sqrt{s} = 13$\,TeV}'',} \textit{ JINST}
  \textbf{ 15} (2020) P10017,
  \href{http://dx.doi.org/10.1088/1748-0221/15/10/P10017}{\doi{10.1088/1748-0221/15/10/P10017}},
  \href{http://www.arXiv.org/abs/2006.10165}{\texttt{arXiv:2006.10165}}.

\bibitem{Khachatryan:2016bia}
\hrefCMSnoop {}{{CMS Collaboration}, ``{The CMS trigger system}'',} \textit{
  JINST} \textbf{ 12} (2017) P01020,
  \href{http://dx.doi.org/10.1088/1748-0221/12/01/P01020}{\doi{10.1088/1748-0221/12/01/P01020}},
\href{http://www.arXiv.org/abs/1609.02366}{\texttt{arXiv:1609.02366}}.

\bibitem{CMS:2024aqx}
\hrefCMSnoop {}{{CMS Collaboration}, ``{Performance of the CMS high-level
  trigger during LHC Run 2}'',} \textit{ JINST} \textbf{ 19} (2024) P11021,
  \href{http://dx.doi.org/10.1088/1748-0221/19/11/P11021}{\doi{10.1088/1748-0221/19/11/P11021}},
  \href{http://www.arXiv.org/abs/2410.17038}{\texttt{arXiv:2410.17038}}.

\bibitem{CMS:2014pgm}
\hrefCMSnoop {}{{CMS Collaboration}, ``{Description and performance of track
  and primary-vertex reconstruction with the CMS tracker}'',} \textit{ JINST}
  \textbf{ 9} (2014) P10009,
  \href{http://dx.doi.org/10.1088/1748-0221/9/10/P10009}{\doi{10.1088/1748-0221/9/10/P10009}},
  \href{http://www.arXiv.org/abs/1405.6569}{\texttt{arXiv:1405.6569}}.

\bibitem{CMS-TDR-15-02}
\href {http://cds.cern.ch/record/2020886}{{CMS Collaboration}, ``Technical
  proposal for the {Phase-II} upgrade of the {Compact Muon Solenoid}'',} CMS
  Technical Proposal CERN-LHCC-2015-010, CMS-TDR-15-02, 2015.

\bibitem{CMS:2017yfk}
\hrefCMSnoop {}{{CMS Collaboration}, ``Particle-flow reconstruction and global
  event description with the {CMS} detector'',} \textit{ JINST} \textbf{ 12}
  (2017) P10003,
  \href{http://dx.doi.org/10.1088/1748-0221/12/10/P10003}{\doi{10.1088/1748-0221/12/10/P10003}},
\href{http://www.arXiv.org/abs/1706.04965}{\texttt{arXiv:1706.04965}}.

\bibitem{CMS:2018rym}
\hrefCMSnoop {}{{CMS Collaboration}, ``{Performance of the CMS muon detector
  and muon reconstruction with proton-proton collisions at $\sqrt{s}=$ 13
  TeV}'',} \textit{ JINST} \textbf{ 13} (2018) P06015,
  \href{http://dx.doi.org/10.1088/1748-0221/13/06/P06015}{\doi{10.1088/1748-0221/13/06/P06015}},
  \href{http://www.arXiv.org/abs/1804.04528}{\texttt{arXiv:1804.04528}}.

\bibitem{Cacciari:2008gp}
\hrefCMSnoop {}{M.~Cacciari, G.~P. Salam, and G.~Soyez, ``{The anti-\kt jet
  clustering algorithm}'',} \textit{ JHEP} \textbf{ 04} (2008) 063,
  \href{http://dx.doi.org/10.1088/1126-6708/2008/04/063}{\doi{10.1088/1126-6708/2008/04/063}},
  \href{http://www.arXiv.org/abs/0802.1189}{\texttt{arXiv:0802.1189}}.

\bibitem{Cacciari:2011ma}
\hrefCMSnoop {}{M.~Cacciari, G.~P. Salam, and G.~Soyez, ``{FastJet user
  manual}'',} \textit{ Eur. Phys. J. C} \textbf{ 72} (2012) 1896,
  \href{http://dx.doi.org/10.1140/epjc/s10052-012-1896-2}{\doi{10.1140/epjc/s10052-012-1896-2}},
\href{http://www.arXiv.org/abs/1111.6097}{\texttt{arXiv:1111.6097}}.

\bibitem{CMS:2016lmd}
\hrefCMSnoop {}{{CMS Collaboration}, ``{Jet energy scale and resolution in the
  CMS experiment in pp collisions at 8 TeV}'',} \textit{ JINST} \textbf{ 12}
  (2017) P02014,
  \href{http://dx.doi.org/10.1088/1748-0221/12/02/P02014}{\doi{10.1088/1748-0221/12/02/P02014}},
\href{http://www.arXiv.org/abs/1607.03663}{\texttt{arXiv:1607.03663}}.

\bibitem{CMS:2019ctu}
\hrefCMSnoop {}{{CMS Collaboration}, ``{Performance of missing transverse
  momentum reconstruction in proton-proton collisions at $\sqrt{s} =$ 13 TeV
  using the CMS detector}'',} \textit{ JINST} \textbf{ 14} (2019) P07004,
  \href{http://dx.doi.org/10.1088/1748-0221/14/07/P07004}{\doi{10.1088/1748-0221/14/07/P07004}},
  \href{http://www.arXiv.org/abs/1903.06078}{\texttt{arXiv:1903.06078}}.

\bibitem{CMS:2021xjt}
\hrefCMSnoop {}{{CMS Collaboration}, ``{Precision luminosity measurement in
  proton-proton collisions at $\sqrt{s} =$ 13 TeV in 2015 and 2016 at CMS}'',}
  \textit{ Eur. Phys. J. C} \textbf{ 81} (2021) 800,
  \href{http://dx.doi.org/10.1140/epjc/s10052-021-09538-2}{\doi{10.1140/epjc/s10052-021-09538-2}},
  \href{http://www.arXiv.org/abs/2104.01927}{\texttt{arXiv:2104.01927}}.

\bibitem{Alwall:2014hca}
J.~Alwall\hrefCMSnoop {}{ { et~al.}, ``{The automated computation of tree-level
  and next-to-leading order differential cross sections, and their matching to
  parton shower simulations}'',} \textit{ JHEP} \textbf{ 07} (2014) 079,
  \href{http://dx.doi.org/10.1007/JHEP07(2014)079}{\doi{10.1007/JHEP07(2014)079}},
\href{http://www.arXiv.org/abs/1405.0301}{\texttt{arXiv:1405.0301}}.

\bibitem{Nason:2004rx}
\hrefCMSnoop {}{P.~Nason, ``A new method for combining {NLO QCD} with shower
  {Monte Carlo} algorithms'',} \textit{ JHEP} \textbf{ 11} (2004) 040,
  \href{http://dx.doi.org/10.1088/1126-6708/2004/11/040}{\doi{10.1088/1126-6708/2004/11/040}},
\href{http://www.arXiv.org/abs/hep-ph/0409146}{\texttt{arXiv:hep-ph/0409146}}.

\bibitem{Frixione:2007vw}
\hrefCMSnoop {}{S.~Frixione, P.~Nason, and C.~Oleari, ``{Matching NLO QCD
  computations with parton shower simulations: the POWHEG method}'',} \textit{
  JHEP} \textbf{ 11} (2007) 070,
  \href{http://dx.doi.org/10.1088/1126-6708/2007/11/070}{\doi{10.1088/1126-6708/2007/11/070}},
\href{http://www.arXiv.org/abs/0709.2092}{\texttt{arXiv:0709.2092}}.

\bibitem{Alioli:2008gx}
\hrefCMSnoop {}{S.~Alioli, P.~Nason, C.~Oleari, and E.~Re, ``{NLO} vector-boson
  production matched with shower in {POWHEG}'',} \textit{ JHEP} \textbf{ 07}
  (2008) 060,
  \href{http://dx.doi.org/10.1088/1126-6708/2008/07/060}{\doi{10.1088/1126-6708/2008/07/060}},
\href{http://www.arXiv.org/abs/0805.4802}{\texttt{arXiv:0805.4802}}.

\bibitem{Alioli:2010xd}
\hrefCMSnoop {}{S.~Alioli, P.~Nason, C.~Oleari, and E.~Re, ``{A general
  framework for implementing NLO calculations in shower Monte Carlo programs:
  the POWHEG BOX}'',} \textit{ JHEP} \textbf{ 06} (2010) 043,
  \href{http://dx.doi.org/10.1007/JHEP06(2010)043}{\doi{10.1007/JHEP06(2010)043}},
\href{http://www.arXiv.org/abs/1002.2581}{\texttt{arXiv:1002.2581}}.

\bibitem{Sjostrand:2014zea}
T.~Sj{\"o}strand\hrefCMSnoop {}{ { et~al.}, ``{An introduction to PYTHIA
  8.2}'',} \textit{ Comput. Phys. Commun.} \textbf{ 191} (2015) 159,
  \href{http://dx.doi.org/10.1016/j.cpc.2015.01.024}{\doi{10.1016/j.cpc.2015.01.024}},
  \href{http://www.arXiv.org/abs/1410.3012}{\texttt{arXiv:1410.3012}}.

\bibitem{Skands:2014pea}
\hrefCMSnoop {}{P.~Skands, S.~Carrazza, and J.~Rojo, ``Tuning {PYTHIA} 8.1: the
  {Monash} 2013 tune'',} \textit{ Eur. Phys. J. C} \textbf{ 74} (2014) 3024,
  \href{http://dx.doi.org/10.1140/epjc/s10052-014-3024-y}{\doi{10.1140/epjc/s10052-014-3024-y}},
  \href{http://www.arXiv.org/abs/1404.5630}{\texttt{arXiv:1404.5630}}.

\bibitem{Sirunyan:2019dfx}
\hrefCMSnoop {}{{CMS Collaboration}, ``{Extraction and validation of a new set
  of CMS \PYTHIA~8 tunes from underlying-event measurements}'',} \textit{ Eur.
  Phys. J. C} \textbf{ 80} (2020) 4,
  \href{http://dx.doi.org/10.1140/epjc/s10052-019-7499-4}{\doi{10.1140/epjc/s10052-019-7499-4}},
\href{http://www.arXiv.org/abs/1903.12179}{\texttt{arXiv:1903.12179}}.

\bibitem{NNPDF:2017mvq}
\hrefCMSnoop {}{{NNPDF} Collaboration, ``{Parton distributions from
  high-precision collider data}'',} \textit{ Eur. Phys. J. C} \textbf{ 77}
  (2017) 663,
  \href{http://dx.doi.org/10.1140/epjc/s10052-017-5199-5}{\doi{10.1140/epjc/s10052-017-5199-5}},
  \href{http://www.arXiv.org/abs/1706.00428}{\texttt{arXiv:1706.00428}}.

\bibitem{Alwall:2008ag}
\hrefCMSnoop {}{J.~Alwall, P.~Schuster, and N.~Toro, ``{Simplified models for a
  first characterization of new physics at the LHC}'',} \textit{ Phys. Rev. D}
  \textbf{ 79} (2009) 075020,
  \href{http://dx.doi.org/10.1103/PhysRevD.79.075020}{\doi{10.1103/PhysRevD.79.075020}},
  \href{http://www.arXiv.org/abs/0810.3921}{\texttt{arXiv:0810.3921}}.

\bibitem{LHCNewPhysicsWorkingGroup:2011mji}
\hrefCMSnoop {}{{LHC New Physics Working Group}, ``{Simplified models for LHC
  new physics searches}'',} \textit{ J. Phys. G} \textbf{ 39} (2012) 105005,
  \href{http://dx.doi.org/10.1088/0954-3899/39/10/105005}{\doi{10.1088/0954-3899/39/10/105005}},
  \href{http://www.arXiv.org/abs/1105.2838}{\texttt{arXiv:1105.2838}}.

\bibitem{CMS:2019zmd}
\hrefCMSnoop {}{{CMS Collaboration}, ``{Search for supersymmetry in
  proton-proton collisions at 13 TeV in final states with jets and missing
  transverse momentum}'',} \textit{ JHEP} \textbf{ 10} (2019) 244,
  \href{http://dx.doi.org/10.1007/JHEP10(2019)244}{\doi{10.1007/JHEP10(2019)244}},
  \href{http://www.arXiv.org/abs/1908.04722}{\texttt{arXiv:1908.04722}}.

\bibitem{CMS:2022sfi}
\hrefCMSnoop {}{{CMS Collaboration}, ``{Search for electroweak production of
  charginos and neutralinos at $\sqrt{s}=13$ TeV in final states containing
  hadronic decays of WW, WZ, or WH and missing transverse momentum}'',}
  \textit{ Phys. Lett. B} \textbf{ 842} (2023) 137460,
  \href{http://dx.doi.org/10.1016/j.physletb.2022.137460}{\doi{10.1016/j.physletb.2022.137460}},
  \href{http://www.arXiv.org/abs/2205.09597}{\texttt{arXiv:2205.09597}}.

\bibitem{Agostinelli:2002hh}
\hrefCMSnoop {}{{GEANT4} Collaboration, ``{\GEANTfour}---a simulation
  toolkit'',} \textit{ Nucl. Instrum. Meth. A} \textbf{ 506} (2003) 250,
  \href{http://dx.doi.org/10.1016/S0168-9002(03)01368-8}{\doi{10.1016/S0168-9002(03)01368-8}}.

\bibitem{Lange:2015sba}
\hrefCMSnoop {}{{CMS Collaboration}, ``{Upgrades for the CMS simulation}'',}
  \textit{ J. Phys. Conf. Ser.} \textbf{ 608} (2015) 012056,
  \href{http://dx.doi.org/10.1088/1742-6596/608/1/012056}{\doi{10.1088/1742-6596/608/1/012056}}.

\bibitem{CMS:2018mlc}
\hrefCMSnoop {}{{CMS Collaboration}, ``{Measurement of the inelastic
  proton-proton cross section at $ \sqrt{s}=13 $ TeV}'',} \textit{ JHEP}
  \textbf{ 07} (2018) 161,
  \href{http://dx.doi.org/10.1007/JHEP07(2018)161}{\doi{10.1007/JHEP07(2018)161}},
  \href{http://www.arXiv.org/abs/1802.02613}{\texttt{arXiv:1802.02613}}.

\bibitem{bn}
\hrefCMSnoop {}{S.~Ioffe and C.~Szegedy, ``{Batch normalization: accelerating
  deep network training by reducing internal covariate shift}'',}
\newblock 2015.
\newblock
  \href{http://www.arXiv.org/abs/1502.03167}{\texttt{arXiv:1502.03167}}.

\bibitem{chollet2015keras}
\hrefCMSnoop {}{F.~Chollet { et~al.}, ``Keras''.} \url{https://keras.io}, 2015.

\bibitem{tensorflow}
\hrefCMSnoop {}{M.~Abadi { et~al.}, ``{TensorFlow: A system for large-scale
  machine learning}'',} 2016.
  \href{http://www.arXiv.org/abs/1605.08695}{\texttt{arXiv:1605.08695}}.

\bibitem{Loshchilov:2017bsp}
\hrefCMSnoop {}{I.~Loshchilov and F.~Hutter, ``{Decoupled weight decay
  regularization}'',}
\newblock 2017.
\newblock
  \href{http://www.arXiv.org/abs/1711.05101}{\texttt{arXiv:1711.05101}}.

\bibitem{CMS:2016mwa}
\hrefCMSnoop {}{{CMS Collaboration}, ``{Measurement of the transverse momentum
  spectra of weak vector bosons produced in proton-proton collisions at $
  \sqrt{s}=8 $ TeV}'',} \textit{ JHEP} \textbf{ 02} (2017) 096,
  \href{http://dx.doi.org/10.1007/JHEP02(2017)096}{\doi{10.1007/JHEP02(2017)096}},
  \href{http://www.arXiv.org/abs/1606.05864}{\texttt{arXiv:1606.05864}}.

\bibitem{CMS:2024lrd}
\hrefCMSnoop {}{{CMS Collaboration}, ``{High-precision measurement of the W
  boson mass with the CMS experiment at the LHC}'',} 2024.
  \href{http://www.arXiv.org/abs/2412.13872}{\texttt{arXiv:2412.13872}}.
  Submitted to \textit{Nature}.

\bibitem{CMS:2006myw}
\href {https://cds.cern.ch/record/922757}{{CMS Collaboration}, ``{CMS} physics:
  Technical design report volume 1: Detector performance and software'',} {CMS
  Technical Design Report} CERN-LHCC-2006-001, CMS-TDR-8-1, 2006.

\bibitem{cmssw}
\href {http://cms-sw.github.io/}{{CMS Collaboration}, ``{CMSSW on Github}''.}
\newblock Accessed: 2023-11-08. \url {http://cms-sw.github.io/}.

\bibitem{CMS:2024twn}
\hrefCMSnoop {}{{CMS Collaboration}, ``{Portable acceleration of CMS computing
  workflows with coprocessors as a service}'',} \textit{ Comput. Softw. Big
  Sci.} \textbf{ 8} (2024) 17,
  \href{http://dx.doi.org/10.1007/s41781-024-00124-1}{\doi{10.1007/s41781-024-00124-1}},
  \href{http://www.arXiv.org/abs/2402.15366}{\texttt{arXiv:2402.15366}}.

\end{thebibliography}\endgroup
\cleardoublepage \appendix\section{The CMS Collaboration \label{app:collab}}\begin{sloppypar}\hyphenpenalty=5000\widowpenalty=500\clubpenalty=5000
\cmsinstitute{Yerevan Physics Institute, Yerevan, Armenia}
{\tolerance=6000
A.~Hayrapetyan, V.~Makarenko\cmsorcid{0000-0002-8406-8605}, A.~Tumasyan\cmsAuthorMark{1}\cmsorcid{0009-0000-0684-6742}
\par}
\cmsinstitute{Institut f\"{u}r Hochenergiephysik, Vienna, Austria}
{\tolerance=6000
W.~Adam\cmsorcid{0000-0001-9099-4341}, J.W.~Andrejkovic, L.~Benato\cmsorcid{0000-0001-5135-7489}, T.~Bergauer\cmsorcid{0000-0002-5786-0293}, M.~Dragicevic\cmsorcid{0000-0003-1967-6783}, C.~Giordano, P.S.~Hussain\cmsorcid{0000-0002-4825-5278}, M.~Jeitler\cmsAuthorMark{2}\cmsorcid{0000-0002-5141-9560}, N.~Krammer\cmsorcid{0000-0002-0548-0985}, A.~Li\cmsorcid{0000-0002-4547-116X}, D.~Liko\cmsorcid{0000-0002-3380-473X}, M.~Matthewman, I.~Mikulec\cmsorcid{0000-0003-0385-2746}, J.~Schieck\cmsAuthorMark{2}\cmsorcid{0000-0002-1058-8093}, R.~Sch\"{o}fbeck\cmsAuthorMark{2}\cmsorcid{0000-0002-2332-8784}, D.~Schwarz\cmsorcid{0000-0002-3821-7331}, M.~Shooshtari, M.~Sonawane\cmsorcid{0000-0003-0510-7010}, W.~Waltenberger\cmsorcid{0000-0002-6215-7228}, C.-E.~Wulz\cmsAuthorMark{2}\cmsorcid{0000-0001-9226-5812}
% ...
\par}
\vskip\cmsinstskip
\dag:~Deceased\\
$^{1}$Also at Yerevan State University, Yerevan, Armenia\\
$^{2}$Also at TU Wien, Vienna, Austria\\
% ...
\end{sloppypar}
\end{document}